\RequirePackage[2020-02-02]{latexrelease}
\documentclass[journal]{IEEEtran}
\IEEEoverridecommandlockouts
\usepackage{amsmath,amssymb,amsfonts}
\usepackage[linesnumbered,ruled]{algorithm2e}
\usepackage{amsmath,algpseudocode}
\usepackage{graphicx}
\usepackage{textcomp}
\usepackage{xcolor}
\usepackage{caption}
\usepackage{subcaption}
\usepackage[inline]{enumitem}

\usepackage{multirow, multicol}
\usepackage{physics}    % \pdv
\usepackage[flushleft]{threeparttable}
\usepackage{afterpage}
\usepackage[nolist]{acronym}

\usepackage{comment}
\usepackage[normalem]{ulem}

\def\BibTeX{{\rm B\kern-.05em{\sc i\kern-.025em b}\kern-.08em
    T\kern-.1667em\lower.7ex\hbox{E}\kern-.125emX}}
    
\usepackage{lipsum}
\usepackage{fancyhdr}

\fancypagestyle{sec}{\lhead{\tmpx}}

\fancypagestyle{arXiv}
{
   \fancyhf{}
   
   \chead{\textcolor{gray}{This paper has been accepted for publication at the \\ IEEE Transactions on Circuits and Systems I - Regular Papers,  2022}}
   
   \cfoot{\vspace{-6mm} \fontsize{=6pt}{10pt}\selectfont  \textcolor{gray}{© 2022 IEEE. Personal use of this material is permitted. Permission from IEEE must be obtained for all other uses, in any current or future media, including reprinting/republishing this material for advertising or promotional purposes, creating new collective works, for resale or redistribution to servers or lists, or reuse of any copyrighted component of this work in other works}\\\fontsize{=10pt}{12pt}\selectfont\thepage}
   
}

\begin{acronym}
\acro{DAS}{Dynamic Audio Sensor}
\acro{DASLP}{Dynamic Audio Sensor Low Power}
\acro{PGA}{Programmable Gain Amplifier}
\acro{AGC}{Automatic Gain Control}
\acro{RMS}{Root Mean Square}
\acro{STD}{Standard Deviation}
\acro{IFR}{instantaneous firing rate}
\acro{SNN}{spiking neural network}
\acro{LIF}{Leaky Integrate-and-Fire}
\acro{FPGA}{field-programmable gate array}
\acro{ASIC}{Application-specific Integrated Circuit}
\acro{LR}{Logistic Regression}
\acro{DRAM}{dynamic random-access memory}
\acro{SRAM}{static random-access memory}
\acro{MAC}{multiply-and-accumulate}
\acro{ISA}{instruction-set architecture}
\acro{SoC}{system-on-chip}
\acro{IoT}{Internet of Things}
\acro{VAD}{voice activity detection}
\acro{LUT}{Look-Up Table}
\end{acronym}

\begin{document}

\title{Spiking Cochlea with System-level Local Automatic Gain Control}

 \author{
     \IEEEauthorblockN{Ilya Kiselev, Chang Gao~\IEEEmembership{Member,~IEEE},
      and  Shih-Chii Liu~\IEEEmembership{Fellow,~IEEE}}\\
    \IEEEauthorblockA{Institute of Neuroinformatics, University of Zurich and ETH Zurich, Zurich, Switzerland
     \\\{kiselev,chang,shih\}@ini.uzh.ch}
    \thanks{Corresponding author: shih@ini.uzh.ch. This work was partially supported by the Swiss National Science Foundation, HEAR-EAR, 200021\_172553.}
 }

\IEEEaftertitletext{\vspace{-2.0\baselineskip}}
\maketitle

\begin{abstract}

Including local automatic gain control (AGC) circuitry into a silicon cochlea design has been challenging because of transistor mismatch and model complexity. To address this, we present an alternative system-level algorithm that implements channel-specific AGC in a silicon spiking cochlea by measuring the output spike activity of individual channels. The bandpass filter gain of a channel is adapted dynamically to the input amplitude so that the average output spike rate stays within a defined range. Because this AGC mechanism only needs counting and adding operations, it can be  implemented at low hardware cost in a future design.
We evaluate the impact of the local AGC algorithm on a classification task where the input signal varies over 32\,dB input range. Two classifier types receiving cochlea spike features 
were tested on a speech versus noise classification task.
The logistic regression classifier 
achieves an average of 6\% improvement and 40.8\% relative improvement in accuracy when the AGC is enabled. The deep neural network classifier shows a similar improvement for the AGC case and achieves a higher mean accuracy of 96\% compared to the best accuracy of 91\% from the logistic regression classifier. 

 \end{abstract}

\section{Introduction}
\thispagestyle{arXiv}
\ac{ASIC} cochlea designs implement circuits that model the filtering properties of the basilar membrane and the rectifying properties of the inner hair cells in biological cochleas~\cite{lyon2010history,fragniereisscc2005,wattscochlea1992}. More recent designs include circuits that generate the asynchronous spiking outputs of the cochlea \cite{Liu2014cochlea,Yang2016cochlea, Wen2009cochlea,Chan2007cochlea,abdalla2005ultrasonic}. 
Few designs include the local automatic gain control (AGC) function of the outer hair cells in the biological cochlea \cite{Wen2009cochlea, drakakis2009agc,hamilton2008active,OdameHaslerTCAS2008} that allow the cochlea to be operated over a large range of sound amplitudes as encountered in natural  environments~\cite{ruggero1992responses}. 
Because of the complexity of local AGC models for VLSI implementation (e.g. ~\cite{lyon2011cascades}) and the effect of transistor mismatch on  equivalent analog ASIC designs, building an ASIC cochlea circuit within a reasonable chip area and with good matching across  multiple filter channels has been challenging. A design with over a hundred filter stages can suffer from mismatch that makes it less usable \cite{Wen2009cochlea} or the resulting size of each channel prohibits the implementation of large number of channels in a reasonable die area~\cite{drakakis2009agc}. 

Discrete hardware implementations such as 
the biomorphic Hopf cochlea system implemented using discrete electronics \cite{van2003active}; and the FPGA implementation \cite{xu2018fpga} of the CAR-FAC AGC model \cite{lyon2011cascades,lyonCARFAC2017} and multi-lite CAR model~\cite{SinghCarLite2019} can get around the transistor mismatch effects but these systems consume more power and are larger than the ASIC designs.

To allow further investigation of low-power audio edge devices that combine spiking cochleas  together with spiking or event-driven processors in natural surroundings, determining the importance of local gain control in using a spike-based sensor for audio tasks is important. A further advantage of local gain control in the sensor is that the average spike rate of the channels is reduced, thereby leading to fewer computes and lower power consumption in the post-processing event-driven or spiking neural network hardware~\cite{davies2018loihi,indiveri2015neuromorphic,kiselev2016event}. 

Example tasks that already use a spiking audio front-end include  azimuthal sound source localization using spike timing information from the  spikes of a binaural hardware cochlea~\cite{xu2019binaural,anumula2018localization,finger2011localization,vanschaik2009localization}, speech recognition \cite{gaoDigit2019, Wu2018spikingsound}, speaker verification,  multi-modal recognition \cite{li2019lip,kiselev2016event} 
and keyword spotting \cite{ceolini2019icassp}. 
A few studies~\cite{uysal2007spikenoise, ZaiLiuReconstruction2015} further demonstrate that with cochlea spikes, 
the accuracy of a word recognition task drops at a slower rate with increasing noise-to-signal input.

Global AGC that is already applied on the microphone outputs (e.g. the AGC on the \ac{DAS} board \cite{Liu2014cochlea}) can help the system to operate in far-field microphone settings. However, it can cause unwanted attenuation of a useful signal in presence of noise outside of the band of interest of a cochlea channel. 

Because of the challenges in  designing on-chip local gain control circuits, we recently proposed a system-level channel-specific AGC mechanism that uses the spiking activity of the individual filter channels on a DAS spiking cochlea 
 to dynamically adapt the local gain of the filters~\cite{kiselev2021agc}. 
 This real-time AGC mechanism does not use floating-point arithmetic; instead, it only needs counters and comparators that can be implemented at low hardware cost on a future cochlea ASIC.

In this work, we give a  detailed description of the AGC mechanism and present extended  response measurements %of the transfer function 
of the cochlea channels in the absence and presence of this AGC. We also show that the AGC leads to a wider linear input range for a DAS spiking cochlea~\cite{Yang2016cochlea}. 

In addition to the results from the logistic regression classifier that uses the spiking outputs in a speech versus noise experiment, we also studied the classification accuracy using a deep neural network (DNN) classifier. 
 These experiments allow us to determine if using spike features from an AGC-enabled cochlea leads to improved accuracy in a speech classification task. 

 Section~\ref{sec:methods_cochlp} describes details of the spiking DAS cochlea architecture, 
 the sensor spike features and the dataset preparation for the speech classification experiments. Section~\ref{sec:cochlp_agc_hw} presents the AGC algorithm and corresponding implementation on the FPGA hardware platform that interfaces to the cochlea. Measurements of the analog and spiking responses of a filter channel with AGC are presented in Section~\ref{sec:cochlp_results} along with classification results in Section~\ref{sec:classifier} to study the impact of AGC in the spiking cochlea for a speech vs noise classification task.

\section{Methods and Material}
\label{sec:methods_cochlp}

\begin{figure}[t]
\centering
\includegraphics[width=0.45\textwidth]{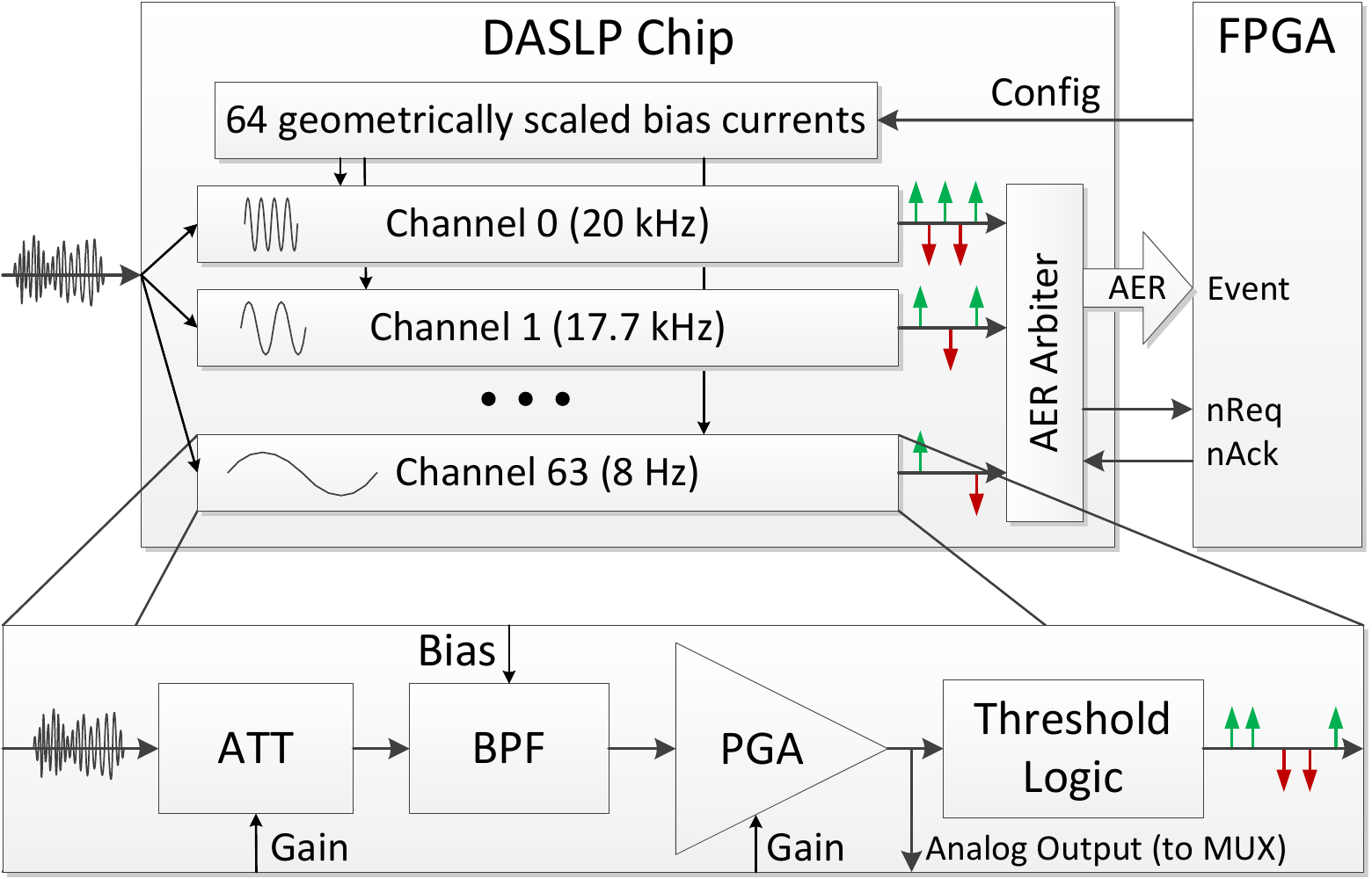}
\caption{Architecture of the spiking DASLP cochlea. A simplified structure diagram of one channel is shown in the bottom callout. The green and red arrows represent ON and OFF spikes.
}
\label{fig:DASLP_arch}
\end{figure}

\subsection{Dynamic Audio Sensor}
\label{sec:methods_daslp}

The ASIC spiking cochlea used in this work is the  low-power binaural \ac{DASLP} 
~\cite{Yang2016cochlea}. Each ear drives a parallel bank of $64$ filters ranging from center frequencies of $20$~Hz to $20$~kHz; and the analog filter core consumes only $55~\mu$W. The center frequency of each filter is generated by the $64$ geometrically-scaled current block in Fig.~\ref{fig:DASLP_arch}. The fourth-order bandpass filter (BPF) design in each channel consists of two cascaded power-efficient second-order source-follower-based BPFs, followed by a spike-generating circuit. 
The spikes are transmitted off-chip using the asynchronous event representation (AER) protocol~\cite{boahen_tcas_2000}. This design has good matching properties of the filter quality factor, $Q$, across the 64 channels and its use was tested on an environmental sound classification task~\cite{ceolini2019audio}.
 
The DASLP test board 
that holds the chip has many  features from a previous DAS board \cite{Liu2014cochlea} but has a more modern USB$3$ interface and a bigger FPGA chip (LATTICE LFE3-70EA-8FN484C)
for the chip control and data readout. The chip-level DASLP AER handshaking signals $\rm{nReq}$ and $\rm{nAck}$ are used to transmit the channel address to this FPGA.  The board is USB powered and interfaces to the java-based jAER software \cite{jaer} which is used for setting the chip biases, recording, and processing of the sensor output. The board also holds a pair of differential $18$-bit ADCs (AD$7691$) that simultaneously sample the input  microphone signals fed into the left and right ears of the cochlea chip. 
The filter output of any channel can be read off chip thereby, allowing measurement of its analog output amplitude. 

Each channel has an individual programmable attenuator (ATT) and a programmable gain amplifier (PGA). 
There are $8$ designed levels of attenuation ranging from $0$ to $-18$ dB and $8$ levels of PGA gain ranging from $18$ dB to $40$ dB~\cite{Yang2016cochlea}. 
Because the chip was not designed to support the fast
switching of the levels needed for the AGC feedback loop, only $4$ attenuation levels ($-6$ dB to $-18$ dB) were used in this work. Loading a new ATT and PGA gain level setting for a channel takes about $0.5$ ms and can only be carried out one channel at a time for this DASLP design.

By combining different ATT and PGA gain levels, 12 overall gain levels were available for the AGC control loop corresponding to a gain range of $0$ dB to $32.5$ dB from average measurements of a set of channels at their center frequency and $Q = 1$. The combinations of the ATT and PGA gains were chosen in a way such that every gain level step change is around $3$~dB. We used 36 channels in this work, corresponding to center frequencies ranging from $56$ Hz to $4$ kHz.

\subsection{Cochlea Spike Features}
\label{sec:SoD_Sampling}
\begin{figure}[t]
\centering
  \includegraphics[width=0.4\textwidth]{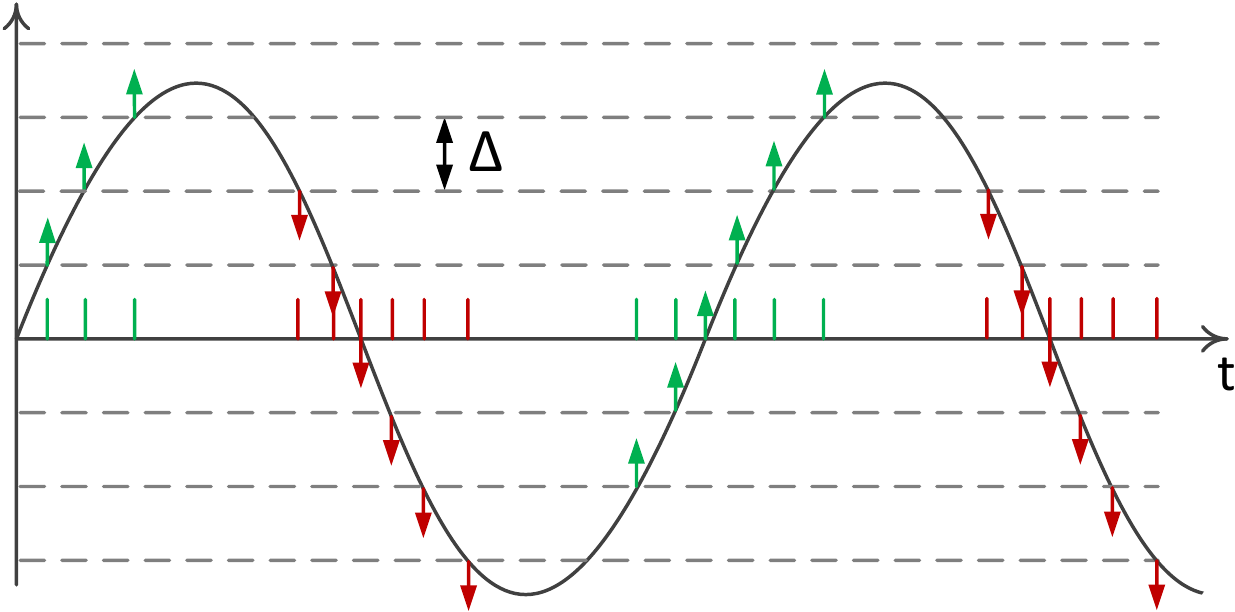}
  \caption{Spike generation using ADM. Arrows represent places where signals cross the threshold, $\Delta$. ON spikes in green. OFF spikes in red.}
  \label{fig:SoD_Spikes}
\end{figure}

The DASLP output spikes are generated using an Asynchronous Delta Modulation (ADM) coding scheme as shown in Fig.~\ref{fig:SoD_Spikes}. ON and OFF events are generated from each channel when its BPF output exceeds a positive or negative threshold relative to the previously encoded level. This scheme preserves information about both the frequency and amplitude of the signal. The number of ON events on the rising slope of the signal encodes the signal amplitude, and the time interval between ON events on two consecutive rising slopes encodes the frequency. 

Different spike features can be extracted from the DAS spikes. These features include constant time spike count, interspike interval histograms and constant count features~\cite{Li2012spid,acharya2018comparison,anumula2018feature}. The features are used by the post-processing algorithm for various audio tasks \cite{gaoDigit2019,Li2012spid,yang2018VAD}. For the local AGC mechanism described in this work, we use constant time bin features which encode the spike count in a fixed time bin.
This time-binning process discards the ordering information of ON and OFF events, leaving only total event counts per time bin per channel. The produced features represent a mixture of both the frequency and amplitude of the signal at the channel. Adjusting the gain of a channel so that only one ON event is generated per period helps to keep the frequency and the amplitude information separated even when time-binning is applied. In this case, the spike count in a time bin encodes the signal frequency while the channel gain  during this time bin encodes the signal amplitude.

\subsection{Dataset Preparation}
\label{sec:dataset}

Speech samples were taken from the TIMIT dataset \cite{garofolo1992timit} and noise samples from both, MS-SNSD \cite{reddy2019scalable} and MUSAN \cite{Snyder2015musan} datasets for the speech versus noise classification experiments. Two hours of speech samples were taken from the TIMIT dataset (one hour from the training set, one hour from the test set) and two hours of noise samples consisting of environmental sounds and music were randomly selected from the MS-SNSD and MUSAN datasets. The noise dataset was randomly split into training and test sets of equal time duration such that each noise sample was included in only the training or test set. 
Each audio file was normalized individually, so that the 
\ac{RMS} amplitude of a signal at the input of the DASLP chip would be equal to the defined amplitude. The audio samples were played to the cochlea through a computer sound card. 
The training set was recorded using five different amplitudes - [$5, 10, 15, 50, 80$]~mV$_{\textrm{RMS}}$ and the test set was recorded using ten different amplitudes - [$2, 2.5, 5, 7, 10, 15, 20, 30, 50, 80$]~mV$_{\textrm{RMS}}$. The maximal gain setting was used for the non-AGC recordings so that there are a few spikes even at the smallest signal amplitude. At this gain level, the input attenuation is set to $-6$~dB, which reduces the signal amplitude at the PGA input by a factor of~$2$.

\section{Event-driven AGC Algorithm and Hardware Implementation}

\label{sec:cochlp_agc_hw}

The gain of global AGC 
commercial chips 
is adjusted according to the amplitude of the measured input. This method treats 
all frequencies 
equally and might drive the useful signal below the detection threshold
in the presence of a high-amplitude noise even when the noise has less power at the  frequency components of the signal.

The AGC control models in previous ASIC  silicon cochleas are based on the output analog amplitude of the channel filter~\cite{drakakis2009agc,hamilton2008active}. A recent model of local gain control in the cochlea is the CAR-FAC model~\cite{lyonCARFAC2017} which uses an additional nonlinear analog filter to adjust the filter parameters. 
To our knowledge, there is little work on using the output spikes to control the local gain of the filters in a cochlea channel.

\subsection{Event-driven AGC Algorithm}
\label{sec:agcAlgorithm}

\begin{figure}[t]
\centering
  \includegraphics[width=0.45\textwidth]{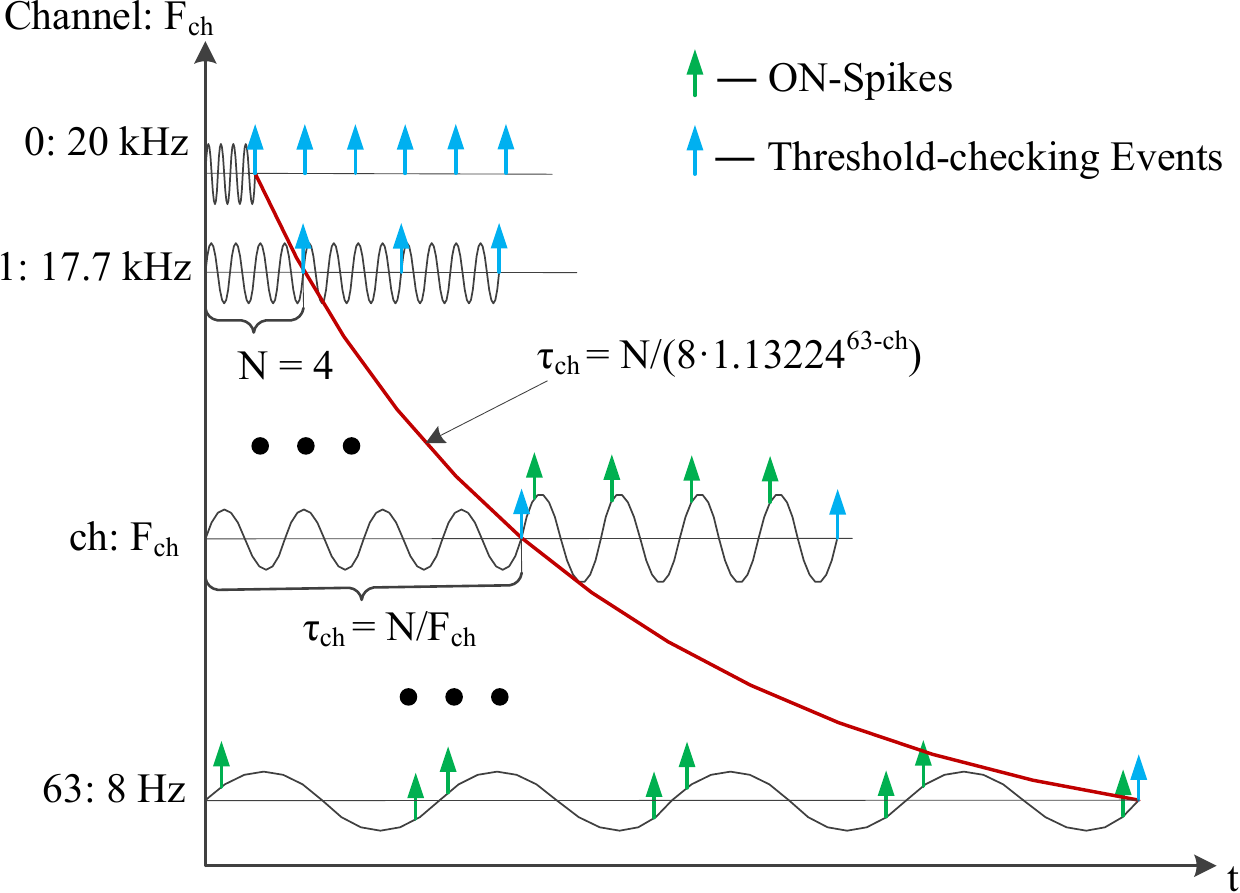}
  \caption{Spike rate averaging time intervals for different channels.}
  \label{fig:AGCspikerateAvg}
\end{figure}

In our gain control mechanism, we use a step-wise feedback controller which is one of the known simpler controllers~\cite{sonneborn1964bang}. This controller is an event-driven controller where fixed-step gain changes are made over time. Using this controller, we 
maintain an average spike rate of each channel within a certain range. The average spike rate, $r_{ch}$, is computed as the number of spikes within an averaging time interval $\tau_{ch}$,  
which is defined individually for each channel based on its center frequency $F_{ch}$. We use $N$ periods of $F_{ch}$ for computing a running estimate of the spike rate $r_{ch}$ of the channel, thus the averaging time interval of channel $ch$, is $\tau_{ch} = N / F_{ch}$. Since the DASLP channels are spaced approximately geometrically within the range of $8$~Hz to $20$~kHz, we use Eq.~\ref{eq:f_ch} to compute 
$F_{ch}$ and hence $\tau_{ch}$ (Eq.~\ref{eq:tau_ch} and red curve in Fig.~\ref{fig:AGCspikerateAvg}),
\begin{equation}
F_{ch} = 8 \cdot sf^{(63-ch)}~~(Hz)
\label{eq:f_ch}
\end{equation}
\begin{equation}
\tau_{ch} = N/(8 \cdot sf^{(63-ch)})~~(s)
\label{eq:tau_ch}
\end{equation}
where $sf$, the scaling factor, is computed by substituting $F_0 = 20$~kHz and $ch=0$ into equation~\ref{eq:f_ch}, $sf \approx 1.13224$.

At the end of $\tau_{ch}$ (denoted by blue arrows in Fig.~\ref{fig:AGCspikerateAvg}), the measured $r_{ch}$ 
is compared to two programmable thresholds: a lower threshold $T_l$ and a upper threshold $T_u$. These thresholds are the same for all channels. When the measured $r_{ch}$ does not fall into the range defined by the two thresholds, the AGC controller changes the channel gain to the next possible value in the direction opposite to the exceeded threshold and starts measuring the spike rate again. In Fig.~\ref{fig:AGCspikerateAvg}, we see an example where the gain of the channel, $ch$, was increased because there were no spikes in the previous averaging time interval $\tau_{ch}$.

A two-threshold control scheme was also used in a previous cochlea design that includes an automatic Q-control (AQC) circuit~\cite{hamilton2008active}. It compares the measured peak analog output amplitude of a filter against two sets of thresholds (Upper/Lower). Whenever the peak value was below the Lower or above the Upper threshold,
a ramp generator circuit generates a current that slowly changes the $Q$ value of the filter. In contrast, our AGC scheme uses the average spike rate measured within an averaging temporal window that is defined by the filter's center frequency instead of the peak value of the output amplitude. 

In order to simplify the implementation of our event-driven AGC algorithm on an FPGA, we use spike counts $sc_{ch}$ over $N$ periods of $F_{ch}$ instead of computing $r_{ch}$
over $N$ periods. 
We also specify the thresholds $T_l$ and $T_u$ as a number of spikes within 
$\tau_{ch}$.
In the experiments described in this work, we set $N = 8$, the lower threshold $T_l=1$ and the upper one $T_u=16$, thereby setting the acceptable range of spike rates as $0.125-2$ spikes per $F_{ch}$ period on average. 

\begin{figure}[t]
\centering
\includegraphics[width=0.45\textwidth]{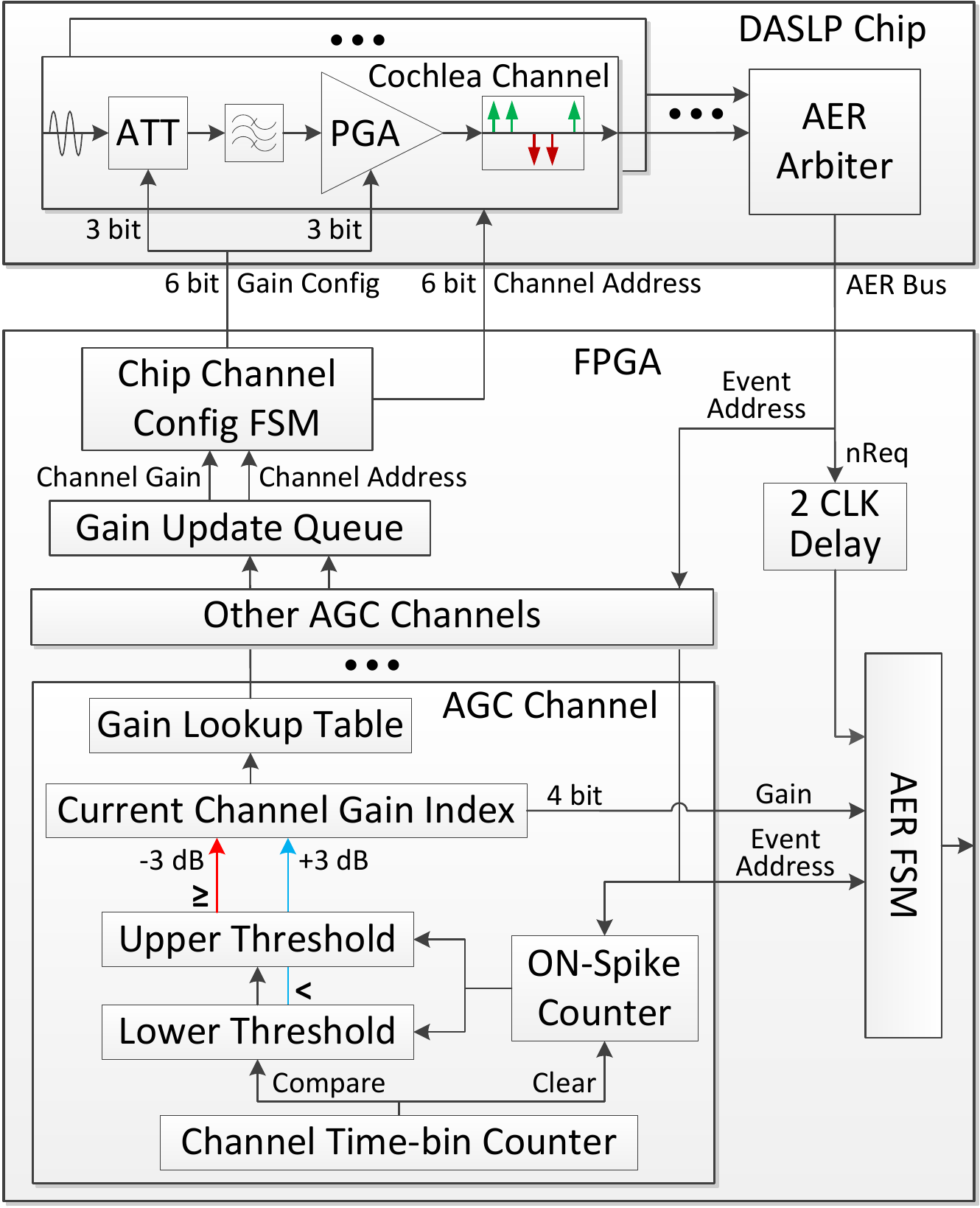}
\caption{Block diagram for FPGA implementation of AGC algorithm.}
\label{fig:AGCchannel-struct}
\end{figure}

The details of the algorithm are described in Algorithm~\ref{alg:AGC_loop}. 
The current channel gain index (0:11) is denoted as $GI[ch]$. The AGC block diagram is shown in Fig.~\ref{fig:AGCchannel-struct}.

\begin{algorithm}[h]
  \caption{AGC control loop for one channel "$ch$"}\label{alg:AGC_loop}
 % \begin{algorithmc}[1]
  \DontPrintSemicolon
  \While{True}{
    \While{$Time\: counter[ch] < \tau_{ch}$}{
      \If{$New\:spike()\; \textbf{and}\; (Spike\: address == ch)$}
        {$Spike\: count[ch] 
       %($sc_{ch}\:
       +\!+$}
      Every $100 \mu$s $Time\: counter[ch] +\!+$\;
    }
    \If {$(Spike\: count[ch] \geq T_u)\; \textbf{and}\; (GI[ch] > 0)$}
      {$GI[ch] -\!-$}
    \If {$(Spike\: count[ch] < T_l)\; \textbf{and}\; (GI[ch] < 11)$}
      {$GI[ch] +\!+$}
    $Current\:gain[ch] = Gain\: lookup\: table[GI[ch]]$\;
    $Time\: counter[ch] = 0$\;
    $Spike\: count[ch] = 0$\;
  }
%\end{algorithmc}
\end{algorithm}
\par

Each spike event from the DASLP cochlea is encoded by 7 bits, of which 6 bits define the  channel address and 1 bit defines the event polarity (Fig.~\ref{fig:SoD_Spikes}).  
The threshold for event generation is the same for all the channels, so the frequency components of the same amplitude would generate the same number of spikes per period at different channels. However, when local AGC is enabled, the gain value of each channel can vary over time. The current gain value can be used for estimating the amplitude of a frequency component at a channel when a spike is generated. Therefore, we embed the current gain setting for the channel that generated an event. Since we have $12$~gain settings, $4$~bits are needed to carry this information. Thus, each spike event carries an additional $4$~bits when the local AGC is enabled. The maximum gain setting ($G_{max}$) corresponds to the channel gain index 11. 

\subsection{FPGA AGC Implementation}
\label{sec:fpga_agc}

The AGC mechanism was implemented on the FPGA of the DASLP board. 
It uses only counters and integer arithmetic and does not require any multipliers, dividers or any other DSP resources on the FPGA (Table~\ref{tab:fpga_resource}).

There are four memory registers associated with each AGC channel: a $12$-bit register for storing the length of the averaging window for a channel in $0.1$\,ms steps;
 $12$-bit register for a counter which represents the time that passed from the beginning of the current averaging window in $0.1$\,ms steps;
 $6$-bit event counter;
 $4$-bit current gain index for a channel.

In addition to these registers, there are two memory bits per channel: A Channel Enable bit (to enable AGC) and an Averaging Window End bit to indicate that the averaging window time has passed.

Thus, $36$ bits of memory are used for one AGC channel. 
There is also a lookup table that translates the $4$-bit channel gain index into a 6-bit gain setting bit pattern. It is shared between all the channels.

By setting the time window counter resolution to $0.1$\,ms and using $12$-bit registers, the length of the averaging time window can be selected in a range from $0.1$\,ms to $409.5$\,ms in steps of $0.1$\,ms. The length of the time window is programmed separately for each channel from the jAER software. The $6$-bit event counter saturates at the value $63$, however it does not overflow -- when it reaches its maximal value, it stops counting and keeps this value until the end of the channel averaging time window. At the end of this window, the event counter is compared to two thresholds. If the spike count is less than the lower threshold, the "gain increase" event is generated, and if the spike count is greater or equal to the upper threshold, the "gain decrease" event is generated.

When any gain update event is generated, it is injected into the AER output FIFO immediately, however updating the gain settings of a channel in the DASLP cochlea chip takes $0.5$ ms, and several gain update requests can occur at the same time or during the time when the system is updating the configuration of the current channel. This would lead to missing gain updates for some channels. To avoid this problem, we implemented a queue for the gain update requests. The queue is implemented as a $128\times12$-bit FIFO, where $6$ bits represent the channel address and the other $6$ bits represent the bit pattern for the new gain setting. The FPGA resources used by the $64$-channel AGC controller are shown in Table~\ref{tab:fpga_resource}.

\begin{table}[!t]
\centering
\caption{Resource utilization of AGC control logic on FPGA LATTICE LFE$3$-$70$EA-$8$FN$484$C.}
\resizebox{0.5\textwidth}{!}{ % RESIZEBOX
\begin{tabular}{|l|c|c|c|c|c|}
\hline
\textbf{}           & \textbf{LUT$4$} & \textbf{LUTRAM} & \textbf{FF} & \textbf{BRAM} ($18$Kb) & \textbf{DSP}   \\ \hline
\textbf{Available}  & $66528$         & $6804$            & $49896$       & $240$                  & $128$  \\ \hline
\textbf{Used}       & $5007$          & $48$              & $3217$        & $1$                    & $0$    \\ \hline
\textbf{Percentage} & $7.5$\%         & $0.7$\%           & $6.4$\%       & $0.4$\%                & $0$\%  \\ \hline
\end{tabular}
}
\label{tab:fpga_resource}
\end{table}

\section{DAS Filter Measurements}
\label{sec:cochlp_results}

We first present measurements of the cochlea filter responses in the presence and absence of the AGC mechanism. Sec.~\ref{ssec:agc_analog_amp} shows the dependence of the frequency response of a filter channel output amplitude for a range of input amplitudes and how it changes when AGC is enabled. Sec.~\ref{ssec:agc_SS_gain_resp} shows the effect of AGC on the steady-state spike responses of the channels.

\subsection{Dependence of Analog Filter Output on Input Amplitude}
\label{ssec:agc_analog_amp}

 AGC is especially beneficial for the 
 DASLP spiking cochlea
that uses a low supply voltage of just $0.5$~V for the analog core. 
To ensure that the transistors circuits operate in the region needed to implement the intended filter transfer function \cite{Yang2016cochlea}, there is a constraint on the 
 \ac{RMS} value of a sine wave input, 
i.e., it should be 
smaller than $\sim$ $5$ mV$_{\textrm{RMS}}$ at the highest gain setting we used (ATT gain=\,-$6$\,dB, PGA gain=$38.5$\,dB).
The BPF drives the spike generating circuit and if the amplitude of the BPF output is too low, spikes will not be produced by the spike generating circuit
even though the incoming frequency is within the passband of the filter. AGC is then useful to amplify the input amplitude to a range so that events will be generated at the channel. In contrast, if the input amplitude is too large, the transistor circuits no longer implement the filter transfer function, therefore AGC is useful in bringing down the input amplitude to the proper operating range. 
\begin{figure}[t]
\centering
  \includegraphics[width=0.48\textwidth]{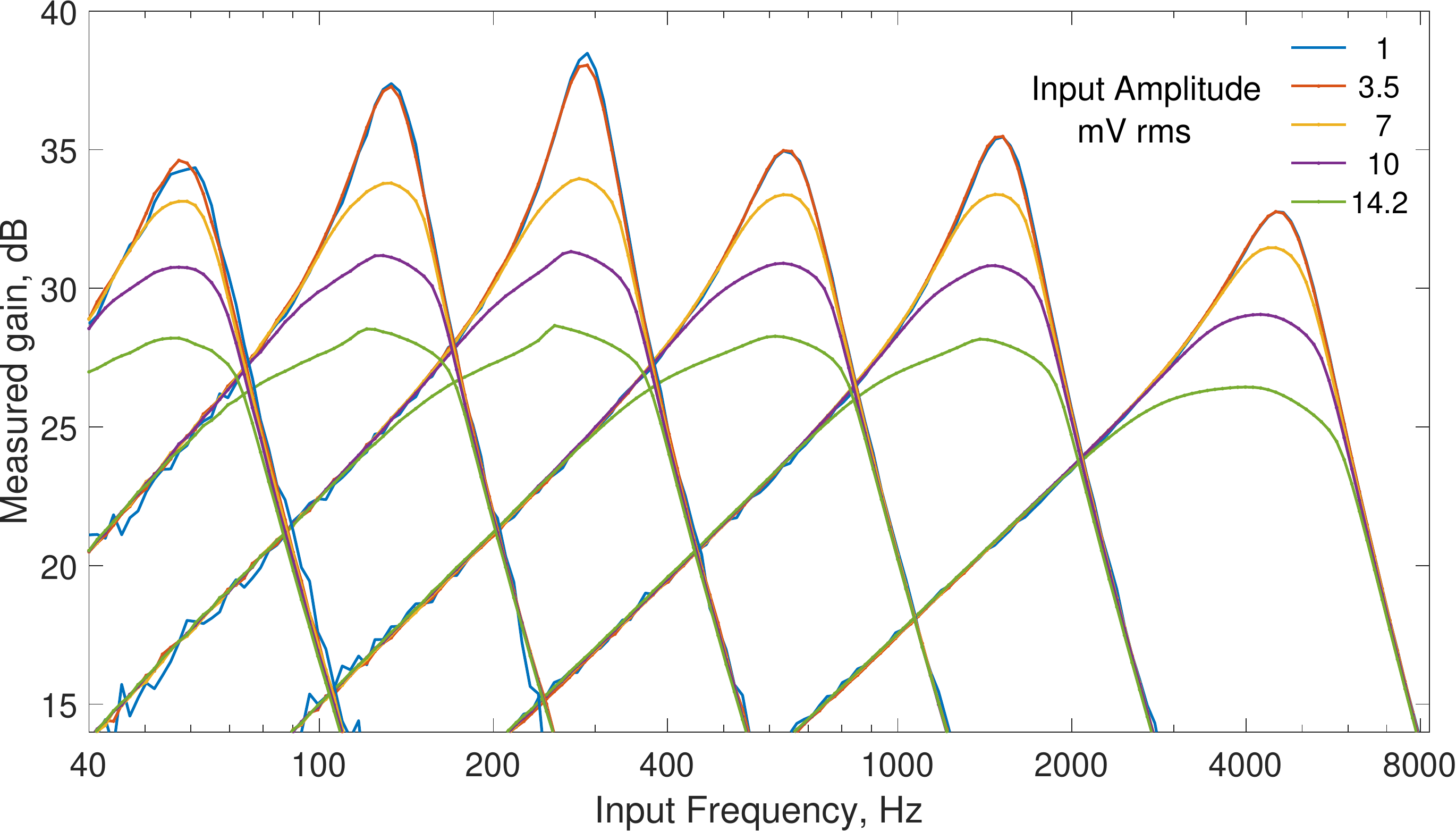}
  \caption{Frequency response of 6 channels 
  with center frequencies of $58$, $133$, $290$, $640$, $1500$ and $4550$~Hz for different input amplitudes. 
  }
  \label{fig:ManyChResponse}
\end{figure}

In order to estimate the input amplitude range of the filters for linear operation in the DASLP 
filters, we measured the frequency response of the channels at different gain settings and input amplitudes.
The input signal to the DASLP
chip and the output bandpass filtered outputs 
are recorded using two differential onboard ADCs 
at $44.1$~kHz sampling rate. The signals of different frequencies were played from a PC to the DASLP cochlea through a sound card. 

The frequency response is computed as a ratio of the output and the input RMS amplitudes. Measurements were done by using the maximum gain setting and $Q\approx 4$ for each channel. However, direct measurement of the output signal amplitude is not possible at low amplitudes due to high noise levels at high PGA gain and high $Q$ value (see Fig.~$18$ in \cite{Yang2016cochlea}). 
The noise RMS amplitude at the highest PGA gain setting %($38.5$~dB) 
and $Q\approx4$ is around $4.5$~mV$_{\textrm{RMS}}$ which is on par with the signal level at low amplitudes. 

In order to account for the noise in the measured output signal, we measured the noise level of each channel at each gain setting and subtracted the corresponding value from the measured signal RMS amplitude following :
\begin{equation}
V_{na} = \sqrt{V_{out}^2 - {V_n(g)}^2}
\label{eq:amplitude}
\end{equation}
where $V_{out}$ is the measured RMS amplitude of the BPF output, $V_n(g)$ is the measured noise level for the gain setting $g$, and $V_{na}$ is the signal amplitude adjusted for noise. 
Thus, the gain of a channel is computed as follows:
\begin{equation}
G = 20*\log_{10}(V_{na}/V_{in})
\label{eq:gain_amplitude}
\end{equation}
where $V_{in}$ is the RMS amplitude of the input signal.

\begin{figure}[!t]
\centering
  \includegraphics[width=0.48\textwidth]{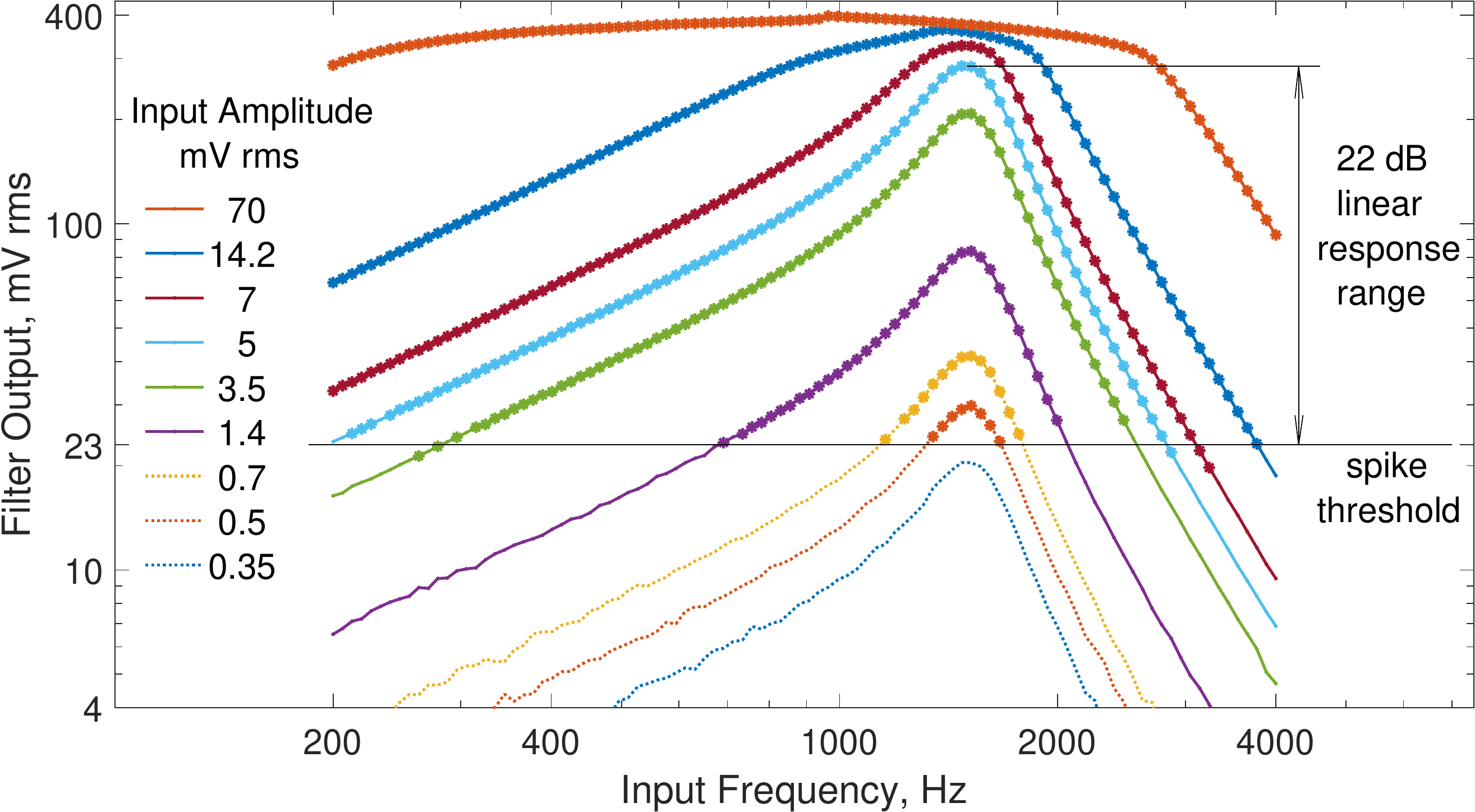}
  \caption{Frequency response of channel $\#23$ at different input amplitudes  with maximum channel gain setting.
  Larger-filled dots indicate points where spikes were obtained.}
  \label{fig:Ch23FreqAmplResponseHG}
\end{figure}

The frequency responses for six channels ($\#48$, $\#42$, $\#36$, $\#30$, $\#23$, $\#13$) are plotted in Fig.~\ref{fig:ManyChResponse}. The curves are measured for five different input
amplitudes, i.e. [$1$, $3.5$, $7$, $10$, $14.2$]~mV$_{\textrm{RMS}}$. The frequency responses of all six channels obtained with the lowest input amplitudes of $1$~mV$_{\textrm{RMS}}$ and $3.5$~mV$_{\textrm{RMS}}$ are well aligned and have the highest measured gain
between $32.5$~dB and $38.5$~dB. For input amplitudes higher than $\approx 7$~mV$_{\textrm{RMS}}$, the shape of the filter curve is distorted
because the transistor circuits no longer implement the intended linear filter transfer function. We see that the peak of the transfer 
curve and the $Q$ factor decreases at every channel for increasing input amplitudes.

Fig.~\ref{fig:Ch23FreqAmplResponseHG} shows the measured output amplitude of channel $\#23$ 
over different input frequencies and amplitudes.
At the maximum gain setting, the output amplitude has to exceed $23$~mV$_{\textrm{RMS}}$ for spikes to be generated at the channel. This minimum output value of $23$~mV$_{\textrm{RMS}}$ corresponds to an input amplitude of $0.4$~mV$_{\textrm{RMS}}$.
The input linear range with reference to the spiking output then varies from $0.4$~mV$_{\textrm{RMS}}$--$5$~mV$_{\textrm{RMS}}$ (light blue curve) which corresponds to 22\,dB.

When the input amplitude increases, the corresponding increase in output amplitude leads to generated spikes  even when the input frequencies are far away from the center frequency. In this case, the response curve becomes non-linear (filter function is distorted) when the input amplitude $>$ $5$~mV$_{\textrm{RMS}}$.

\begin{figure}[t]
\begin{subfigure}{0.5\textwidth}{
\centering
  \includegraphics[width=0.8\textwidth]{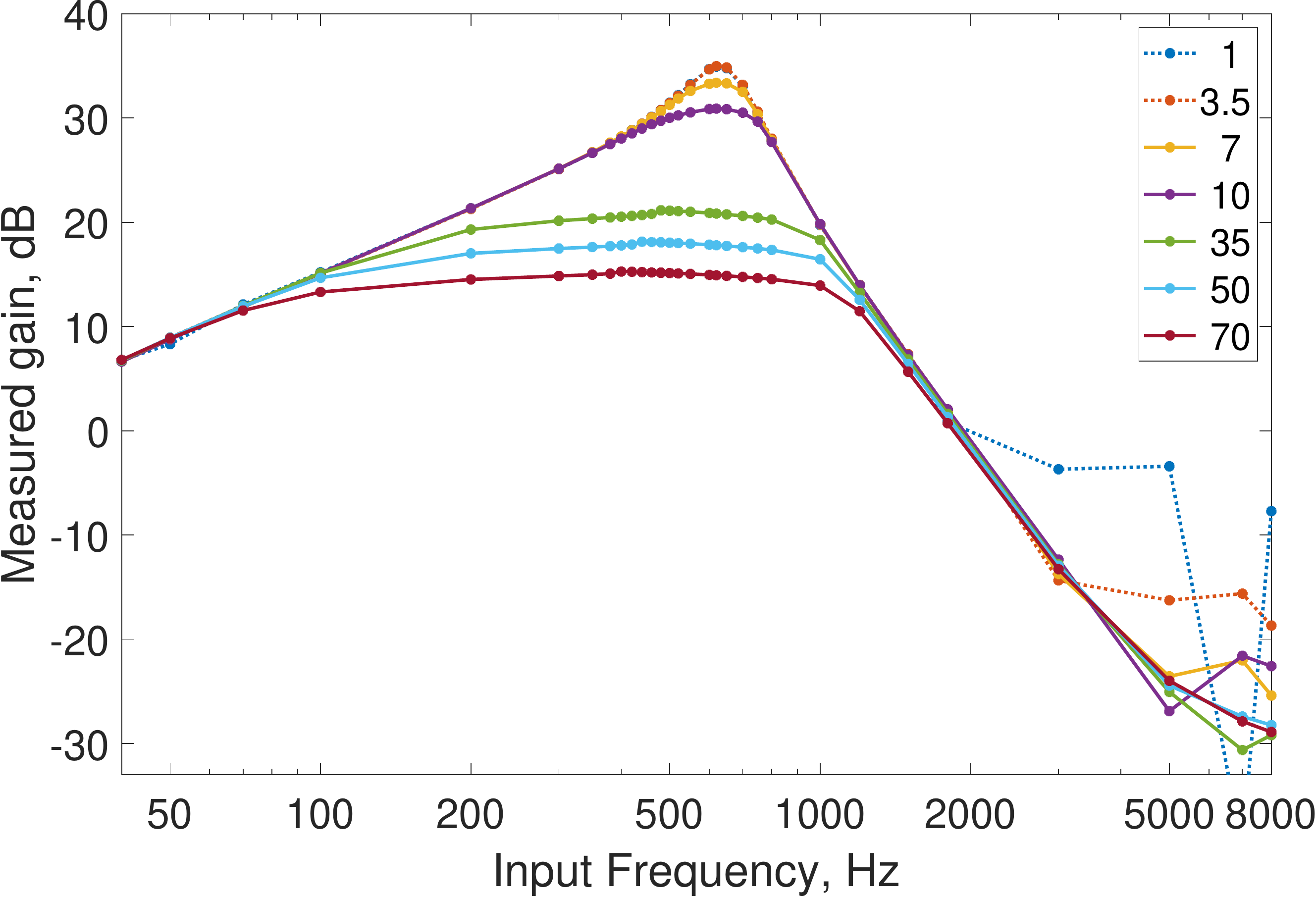}
  \caption{AGC is disabled. }
  \label{fig:FreqGainResponse30}}
  \end{subfigure}
  
\begin{subfigure}{0.5\textwidth}{
\centering
  \includegraphics[width=0.8\textwidth]{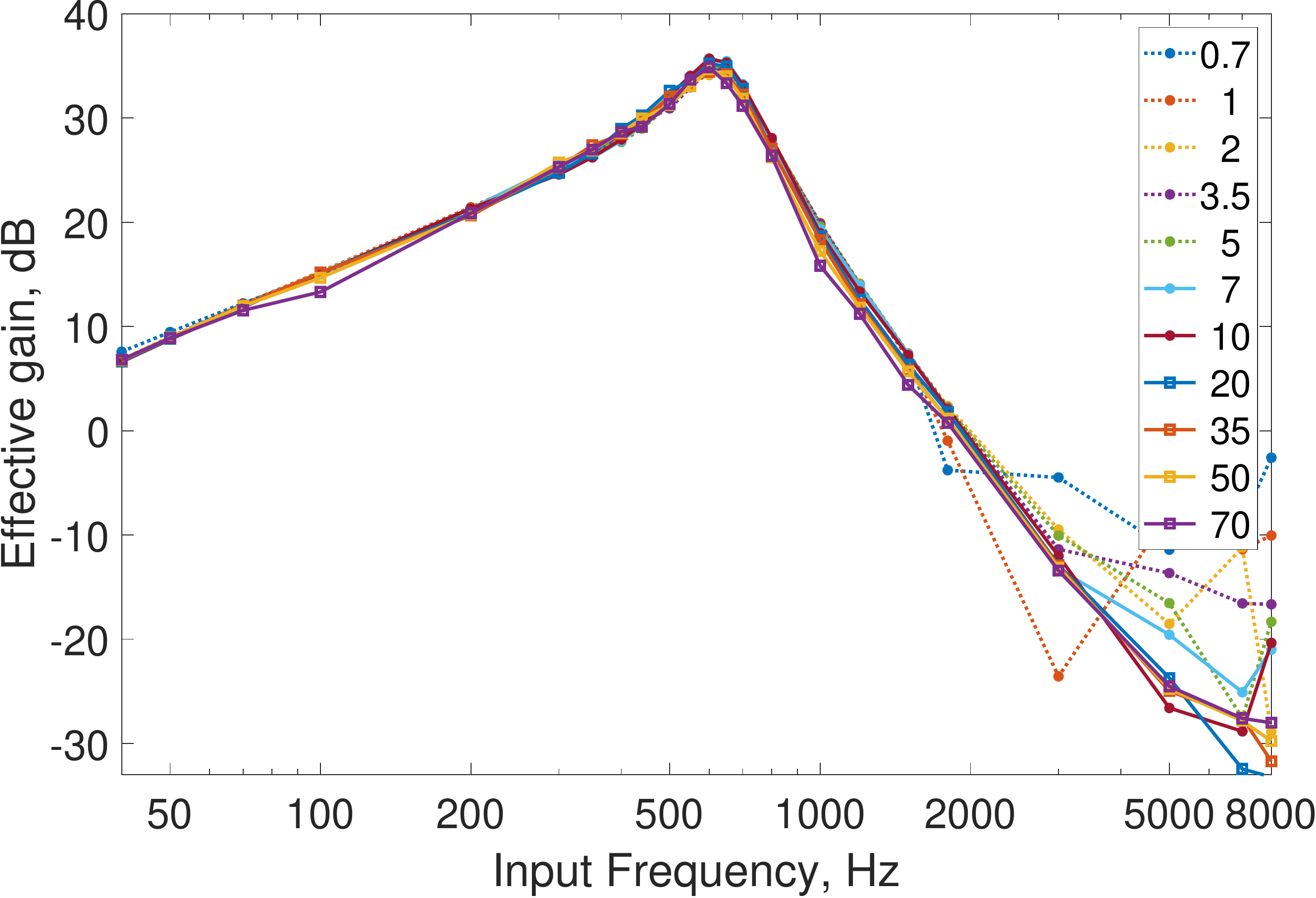}
  \caption{ AGC is enabled.}
  \label{fig:AGCFreqGainResponse30}}
  \end{subfigure}
  
    \caption{Frequency response of channel $\#30$ for different input RMS amplitudes when (a) AGC is disabled and the highest gain setting is used. (b) shows the effective gain responses when AGC is enabled. 
  Input amplitude in legend is in mV$_{\textrm{RMS}}$.}
  \label{fig:freqRespch30}
\end{figure}

We measured the frequency response of another channel ($\# 30$) for different input amplitudes. 
The curves in Fig.~\ref{fig:freqRespch30}(a) were obtained by measuring the steady-state output amplitudes and corresponding gain responses for each input amplitude. The results show that the frequency response for the lowest amplitude has the highest gain and $Q$ factor and as the amplitude increases, the filter function is distorted
while the gain drops. We compute the effective gain $G_e(f)$, when AGC is enabled, i.e.
$G_e(f) = G(f) + C(f)$, where   
$G(f)$ is the channel gain (Eq.~\ref{eq:gain_amplitude}), and $C(f)$ is the compression factor that depends on the gain setting  
introduced by the AGC at the frequency $f$.
The effective gain responses in
Fig.~\ref{fig:freqRespch30}(b) shows that these curves for the different input amplitudes are now well-aligned.

\subsection{AGC Steady-state Spike Response Measurements}
\label{ssec:agc_SS_gain_resp}

\begin{figure}[t]
\centering
\includegraphics[width=0.45\textwidth]{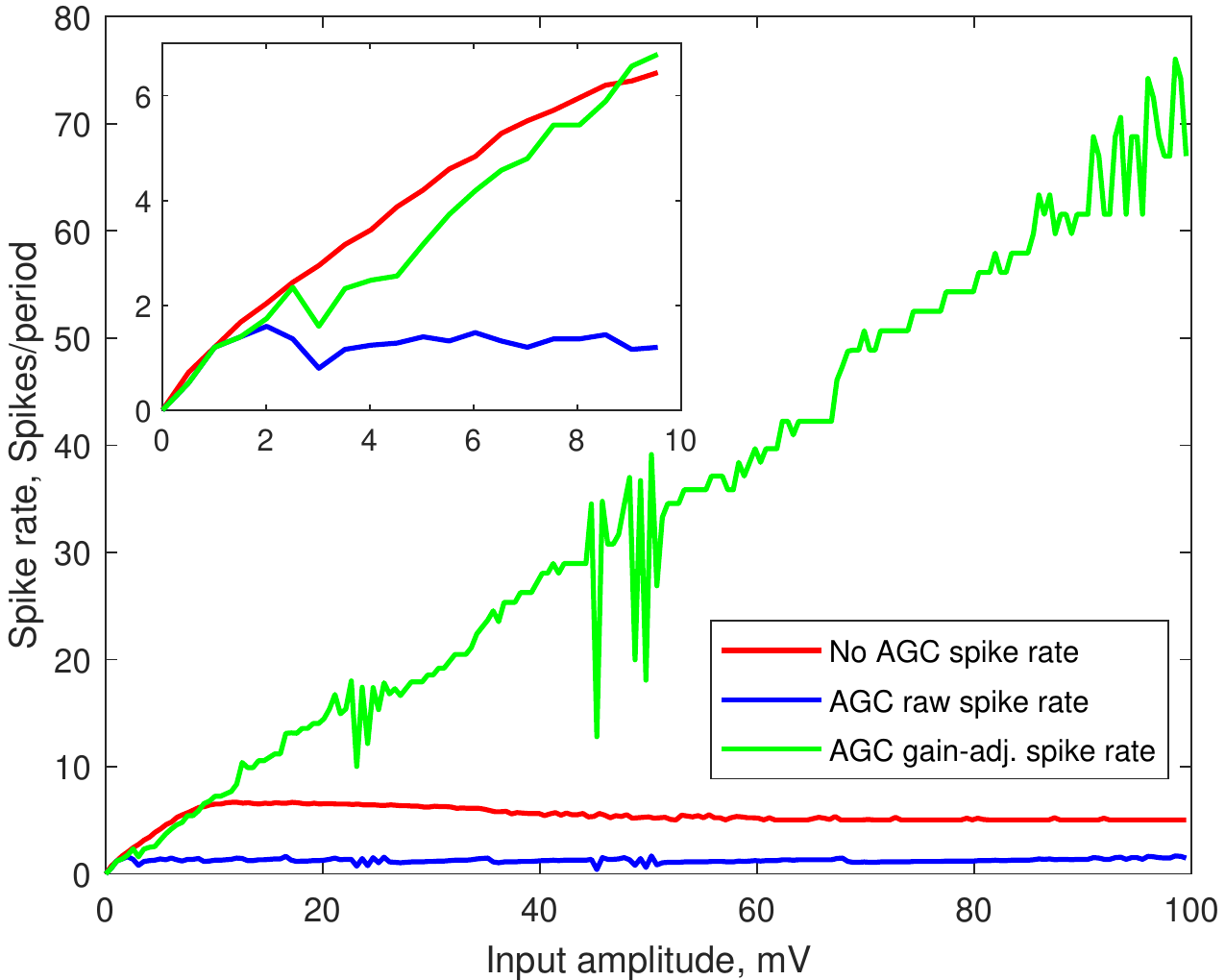}%ch30_AGCnoAGC_spikeRate.png}
\caption{Frequency-normalized spike rate of channel $\#30$ ON-spikes without AGC (red curve) and with AGC (blue curve).
The gain-adjusted frequency-normalized spike rate with AGC is also shown (green curve). Inset in top left corner shows a zoomed-in view for input from $0-10$~mV.}
\label{fig:agc-spikerates}
\end{figure}

The steady-state spike responses in Fig.~\ref{fig:agc-spikerates} show the frequency-normalized spike rates (spikes per input signal period) measured from channel $\#30$ for an input sine wave ($500$~Hz) with amplitudes ranging from $0-100$~mV$_{\textrm{RMS}}$. 
Without AGC (red curve), the spike rate first increases with the input amplitude, but then quickly saturates when the filter goes out of small-signal operation. So, it is not possible to reconstruct the amplitude of the input signal based on the spikes from one channel. 
With AGC (blue curve), the spike rate stays approximately constant over $40$~dB range of input amplitude ($1-100$~mV$_{\textrm{RMS}}$). Because we transmit with the event, the current gain index which corresponds to the gain value of the channel, $G_{ch}$, we can use this information to estimate the current amplitude of the input signal. The higher the channel gain, the lower the input signal amplitude $A_{in}$ needed to produce the same spike rate at the channel. 
We estimated the input amplitude
by dividing the spike rate by the gain of the channel (Eq.~\ref{eq:gain_amplitude}).
\begin{equation}
\hat{A}_{in} \propto s_{ch} / 10^{G_{ch}/20} = s_{ch} \cdot 10^{-G_{ch}/20}
\end{equation}
where $\hat{A}_{in}$ is the estimated input amplitude and  $s_{ch}$ is the frequency-normalized spike rate.

For the non-AGC plot in Fig.~\ref{fig:agc-spikerates}, the channel gain is set to the maximum gain setting,
so in order to compare AGC and non-AGC cases, we use a coefficient $10^{G_{max}/20}$ to equate the estimated spike rate for non-AGC and AGC cases at small input amplitudes, when the AGC also uses the highest gain:
\begin{equation}
r_{ga} = 10^{G_{max}/20} \cdot s_{ch} \cdot 10^{-G_{ch}/20} = s_{ch} \cdot 10^{(G_{max}-G_{ch})/20}
\end{equation}
where $r_{ga}$ is the gain-adjusted spike rate.
The results show that $r_{ga}$ (green curve) is linearly proportional to the input amplitude in a wide range ($40$~dB) of input amplitudes. The spiky glitches in the response are due to transient processes during switching between some gain setting values.

With the help of the AGC, the spike rate of the channel is compressed by $4$x over a wide range of input amplitudes, from an average spike rate of 4.35 kSpikes/sec in the non-AGC case, to 1.09 kSpikes/sec in the AGC case.

\subsection{AGC Transient Measurements to Speech}
\label{sec:gain_resp}

\begin{figure}[t]
\centering
\includegraphics[width=0.487 \textwidth]{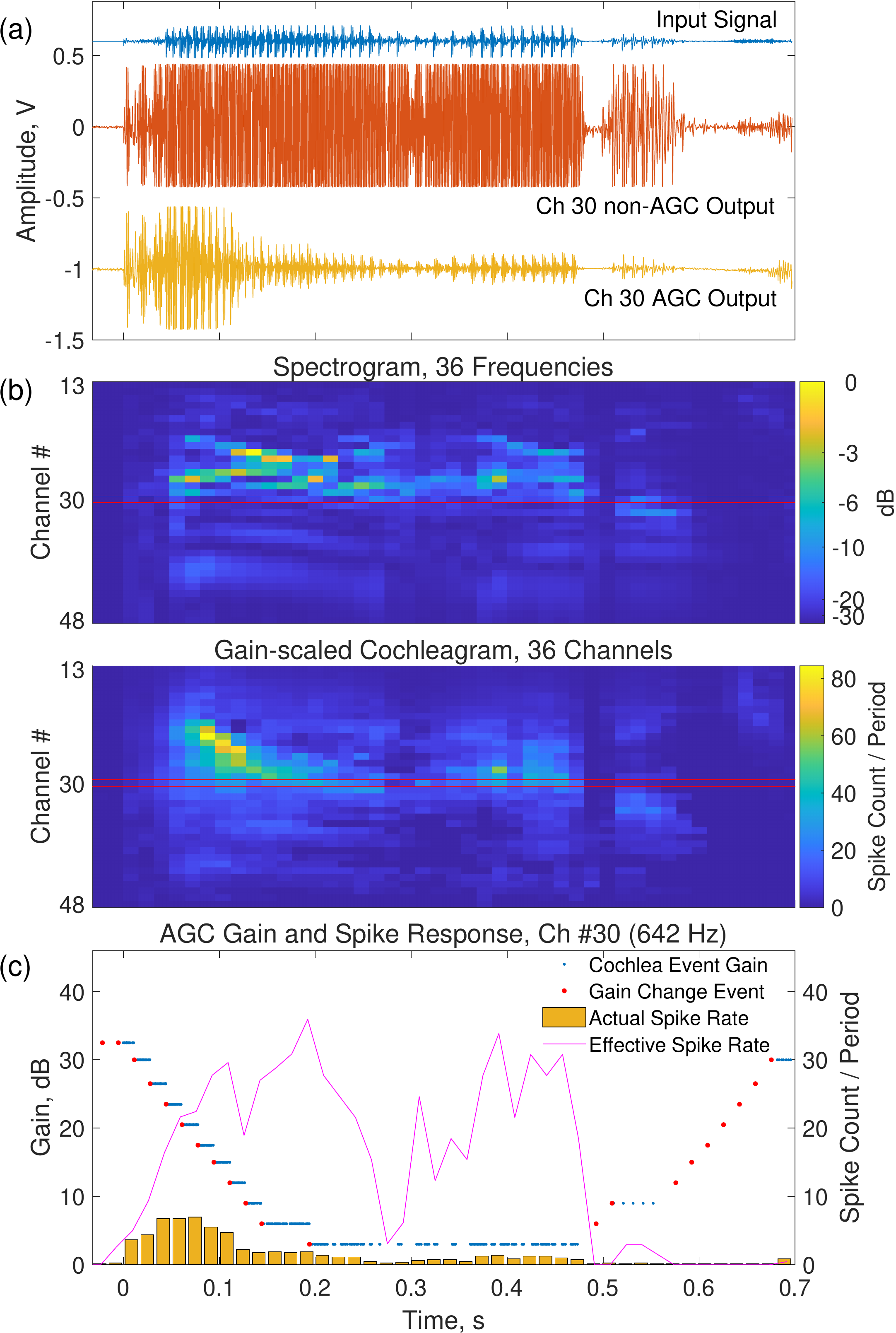}
\caption{AGC response of channel $\#30$~($642$ Hz) to the speech sample "Power out" (male voice) from TIMIT. (a) Waveforms for input signal (blue) and channel output in non-AGC (red) and AGC (yellow) cases. Curves are at an offset for visibility. (b) Spectrogram 
 and gain-scaled cochleagram for $36$  channels with equal center frequencies. Cochlea spike counts are scaled according to the current channel gain within each time bin. (c) AGC gain change events and spike responses. Details in text. }
\label{fig:agc_response_ch30}
\end{figure}

We looked at the effect of the AGC 
to dynamically changing input amplitudes in natural sounds such as speech. We plotted the input signal and the analog output of the bandpass filter of one channel ($\#30$, $642$~Hz) in Fig.~\ref{fig:agc_response_ch30}(a) in response to a speech sample. The speech spectrogram and the gain-scaled cochleagram for $36$ out of the $64$ channels are plotted in Fig.~\ref{fig:agc_response_ch30}(b). The channel spike count response and the gain change events in response to this signal are plotted in Fig.~\ref{fig:agc_response_ch30}(c). 
The blue dots indicate the cochlea ON-spikes at the selected channel, their ordinate represents the current gain of the channel at the time when the event was generated. The gain change events (red dots) indicate gain setting changes when the spike count in the time window has exceeded one of the $2$ thresholds. The ordinate of the dot shows a new gain value for the selected channel. The yellow bars show the average spike count per averaging window length.
The magenta line shows the effective spike rate or the spike rate 
needed in the non-AGC case to represent the channel's analog output signal at the highest gain setting, assuming no signal clipping.

\begin{figure}[t]
\begin{subfigure}{0.5\textwidth}{
\centering
\includegraphics[width=0.75\textwidth]{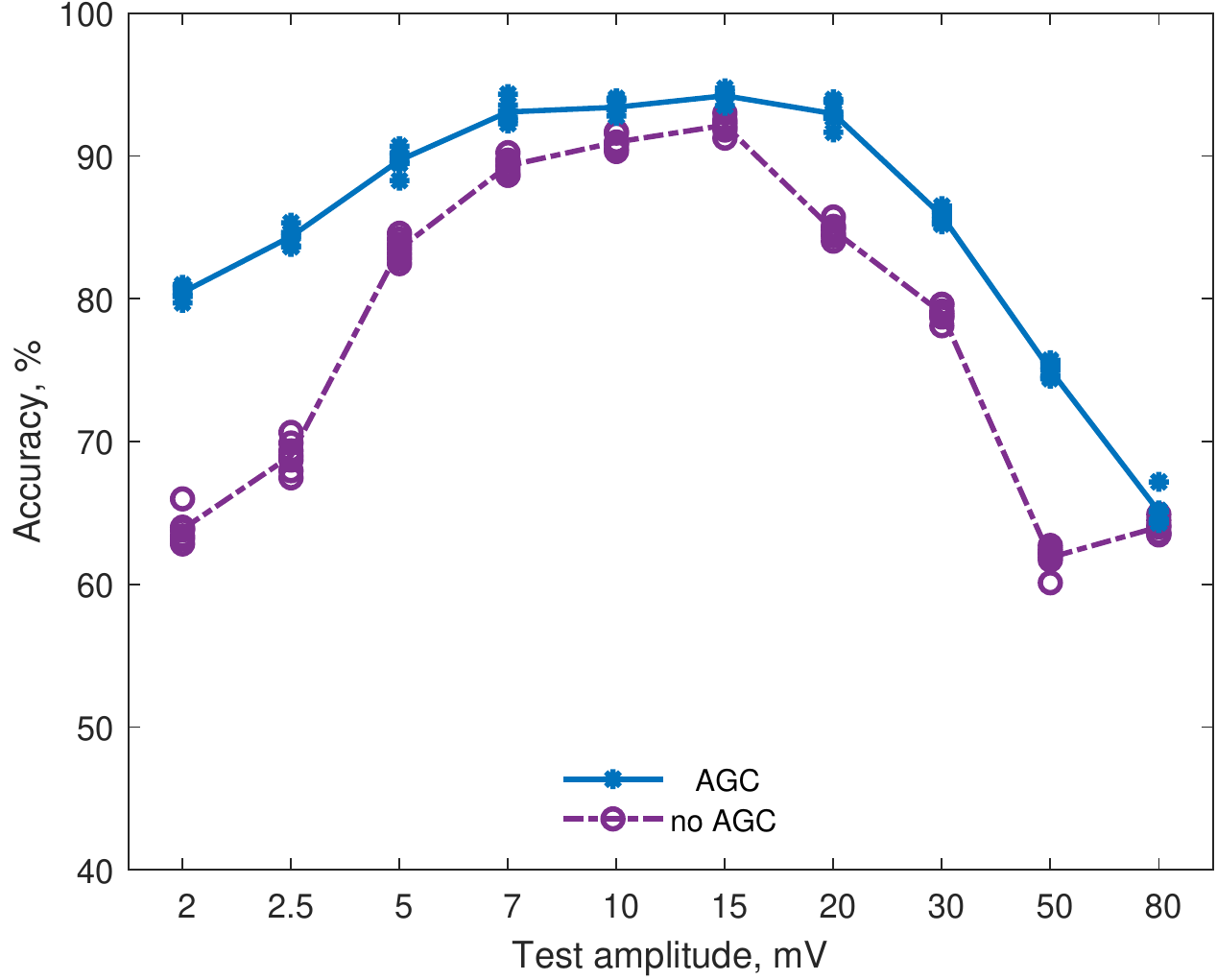}
\caption{LR classifier trained on $15$~mV$_{\textrm{RMS}}$ input.
}
\label{fig:vad15}}
\end{subfigure}

\begin{subfigure}{0.5\textwidth}{
\centering
\includegraphics[width=0.75\textwidth]{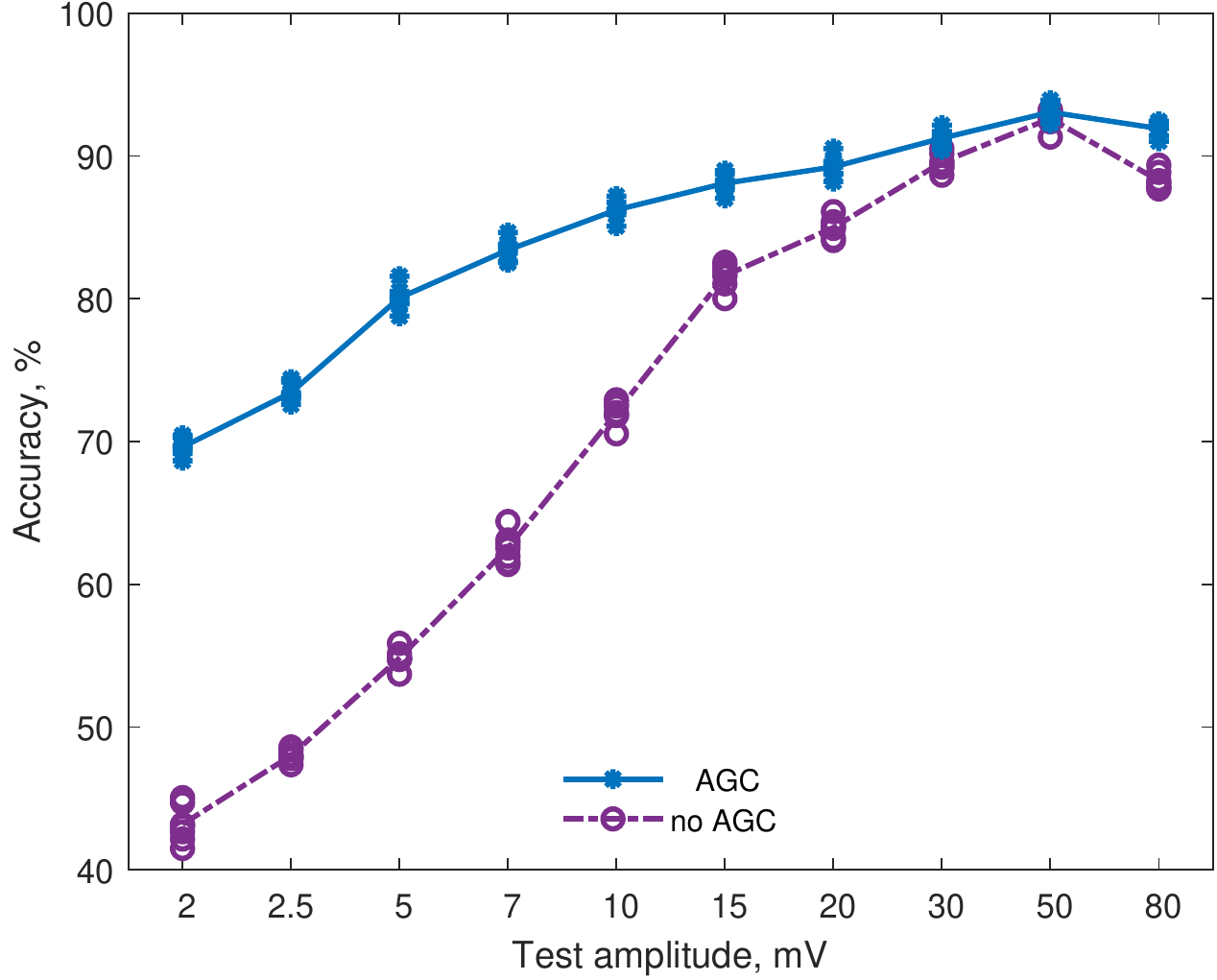}
\caption{LR classifier trained on $50$~mV$_{\textrm{RMS}}$ input.
}
\label{fig:vad50}}
\end{subfigure}

\begin{subfigure}{0.5\textwidth}{
\centering
\includegraphics[width=0.75\textwidth]{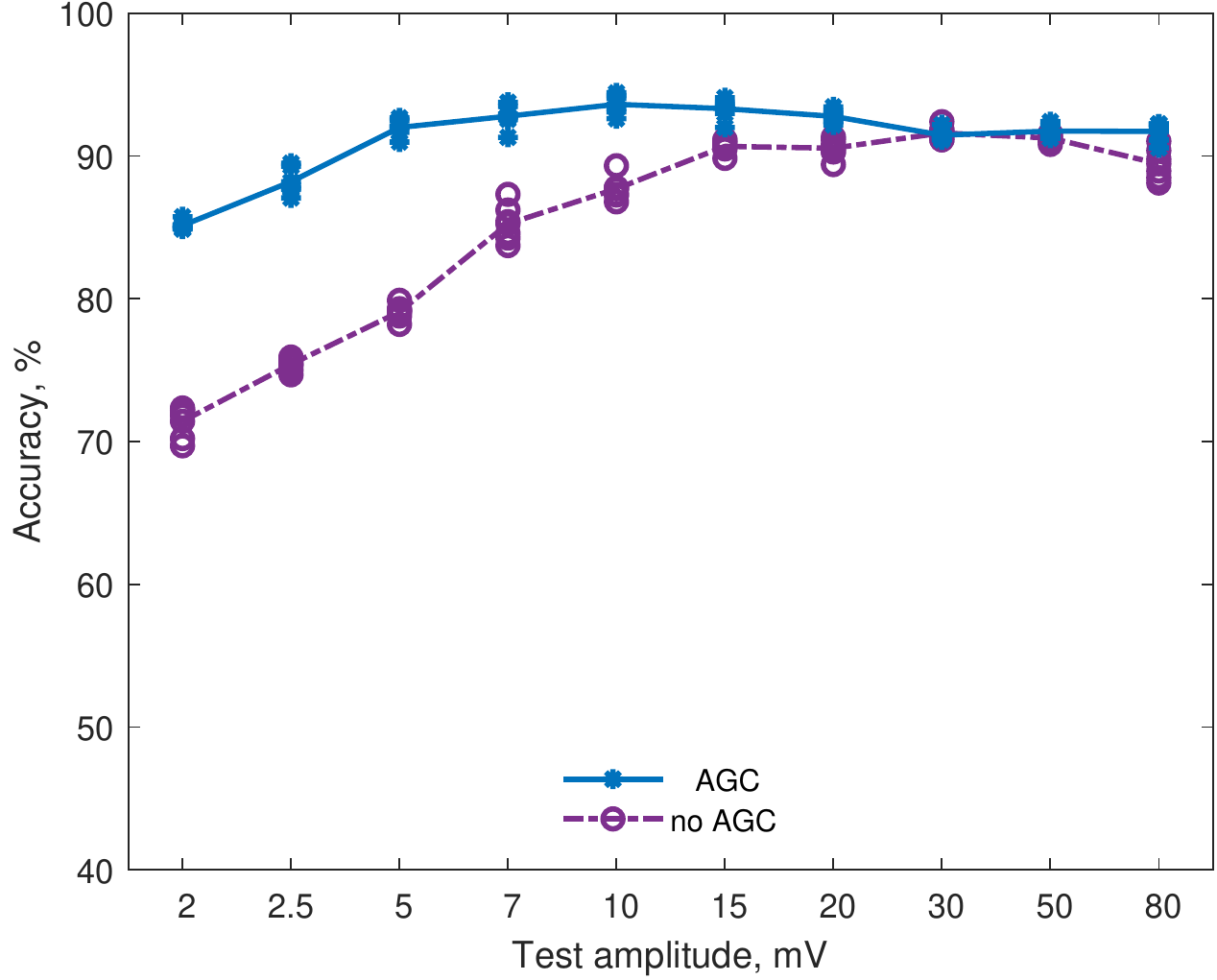}
\caption{LR classifier trained on all input amplitudes.
}
\label{fig:vadall}}
\end{subfigure}
\caption{Test accuracy of speech vs noise LR classifiers trained on recordings at either (a) $15$~mV$_{\textrm{RMS}}$, (b) $50$~mV$_{\textrm{RMS}}$, or (c) all input amplitudes. Testing was done on amplitudes from $2$ to $80$~mV$_{\textrm{RMS}}$.
}
\label{fig:vad_lr}
\end{figure}

This plot also shows that the channel produces no spikes at the beginning because there is no input signal, therefore the gain stays at the highest setting.
When the speech sample starts, the output amplitude is initially high because of the high gain. Over a period of about $150$~ms, the channel gain 
decreases in steps until the spike count (yellow bars in Fig.~\ref{fig:agc_response_ch30}(c)) 
drops below two spikes per period. The spike count represents the number of cochlea spikes per period of channel $\#30$'s center frequency averaged over a window of $16.6$\,ms. The width of the bars is equal to the length of the averaging time window. The gain setting stays constant at $3$~dB during the period of $0.2-0.5$~s, when the spike rate fits within the desirable range, and then gradually increases close to the maximal value during the period of $0.56-0.68$~s, when no spikes are produced at this channel.

\section{Classification Results}
\label{sec:classifier}

\subsection{Input Features}

We evaluate the features generated from only $36$ channels of the spiking cochlea outputs in a speech versus noise classification task. We excluded $12$ high-frequency channels because too many gain updates will be needed for these channels and it would increase the update time for the remaining channels. We excluded the $16$ lowest frequency channels because there were few events generated from the dataset.
Similar to the features described in~\cite{Li2012spid}, we use inter-spike interval histogram and spike count features computed within a $400$~ms frame without overlap. We also include the average AGC gain setting of each channel as part of the input feature. The resulting 152-dimensional feature vectors consist of the $80$-bin histogram of inter-spike intervals, event count of $36$ channels, and $36$ values representing the average gain of each channel. Inter-spike intervals for all channels are computed separately and then combined into one histogram. Inter-spike intervals that are greater than $150$~ms are excluded before the histogram is computed. 
For the "non-AGC" case, the $36$ values representing the average gain of each channel are set to a constant value.

\begin{figure}[!t]
\centering
\includegraphics[width=0.4\textwidth]{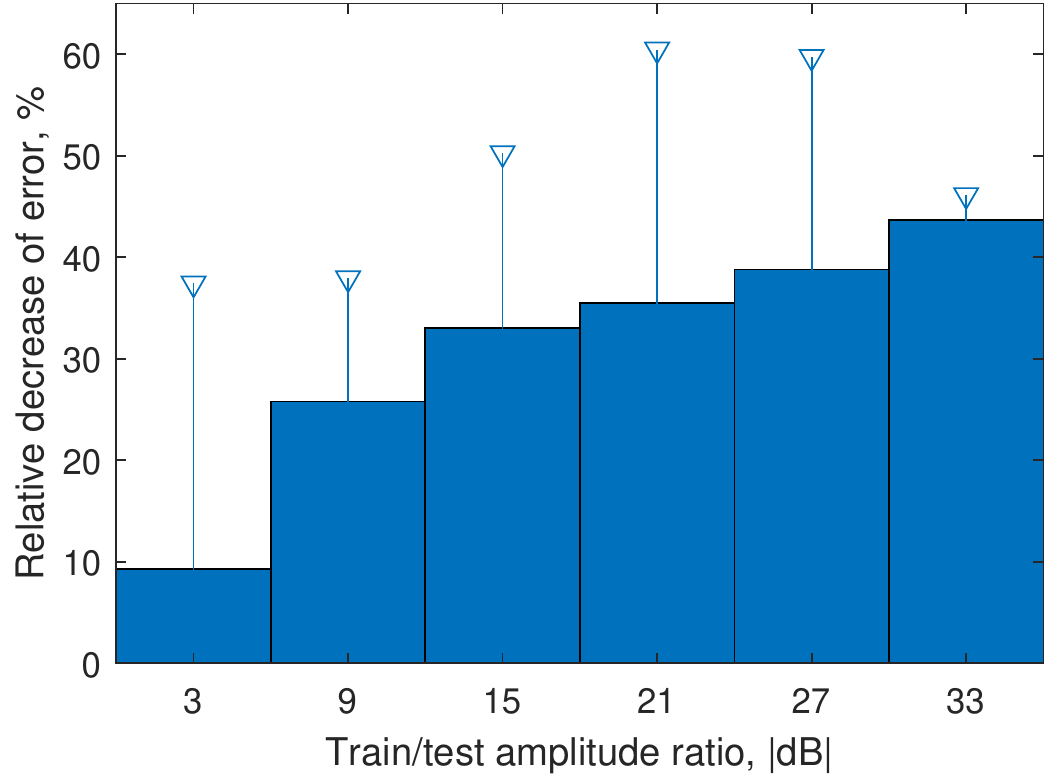}
\caption{Average relative  improvement of error rate (or increase of accuracy) of speech vs noise classifiers trained at different input amplitudes for the AGC vs non-AGC cases and tested over the whole range of input test amplitudes.
}
\label{fig:vad_error_decrease}
\end{figure}

\subsection{Speech versus Noise Classification Task}
\label{sec:classification}

We tested the cochlea features using two classifier methods: A logistic regression (LR) classifier using FP64 precision and a Deep Neural Network (DNN) classifier using FP32 precision. 
 The  classifiers were trained on the spike responses of the cochlea operated in the AGC and the non-AGC mode using the dataset described in Section~\ref{sec:dataset}. Training was carried out for 2 cases. In the first case, we trained the classifier on a single input amplitude and tested across the range of amplitudes ($2$ to $80$~mV$_{\textrm{RMS}}$) in the test set. In the second case, we trained the classifier on all input amplitudes in the training set and then tested on the entire range of test amplitudes. Note that the training set has recordings for 5 input amplitudes while the test set has recordings for 10 amplitudes (including the 5 amplitudes in the training set).

\begin{figure}[t]
\begin{subfigure}{0.5\textwidth}{
\centering
\includegraphics[width=0.8\textwidth]{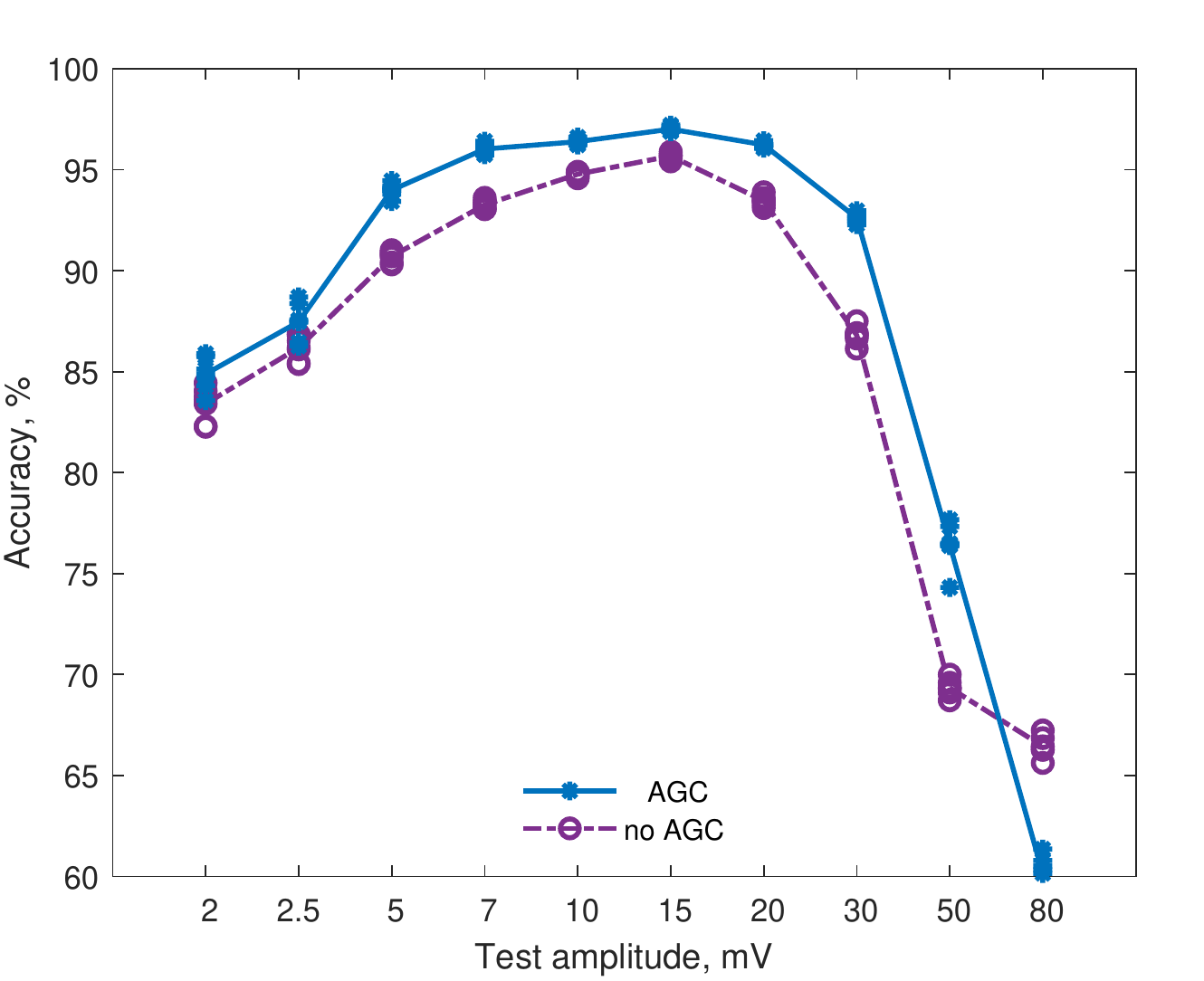}
\caption{DNN trained on $15$~mV$_{\textrm{RMS}}$ input in the training set.
}
\label{fig:exp_feat}}
\end{subfigure}

\begin{subfigure}{0.5\textwidth}{
\centering
\includegraphics[width=0.8\textwidth]{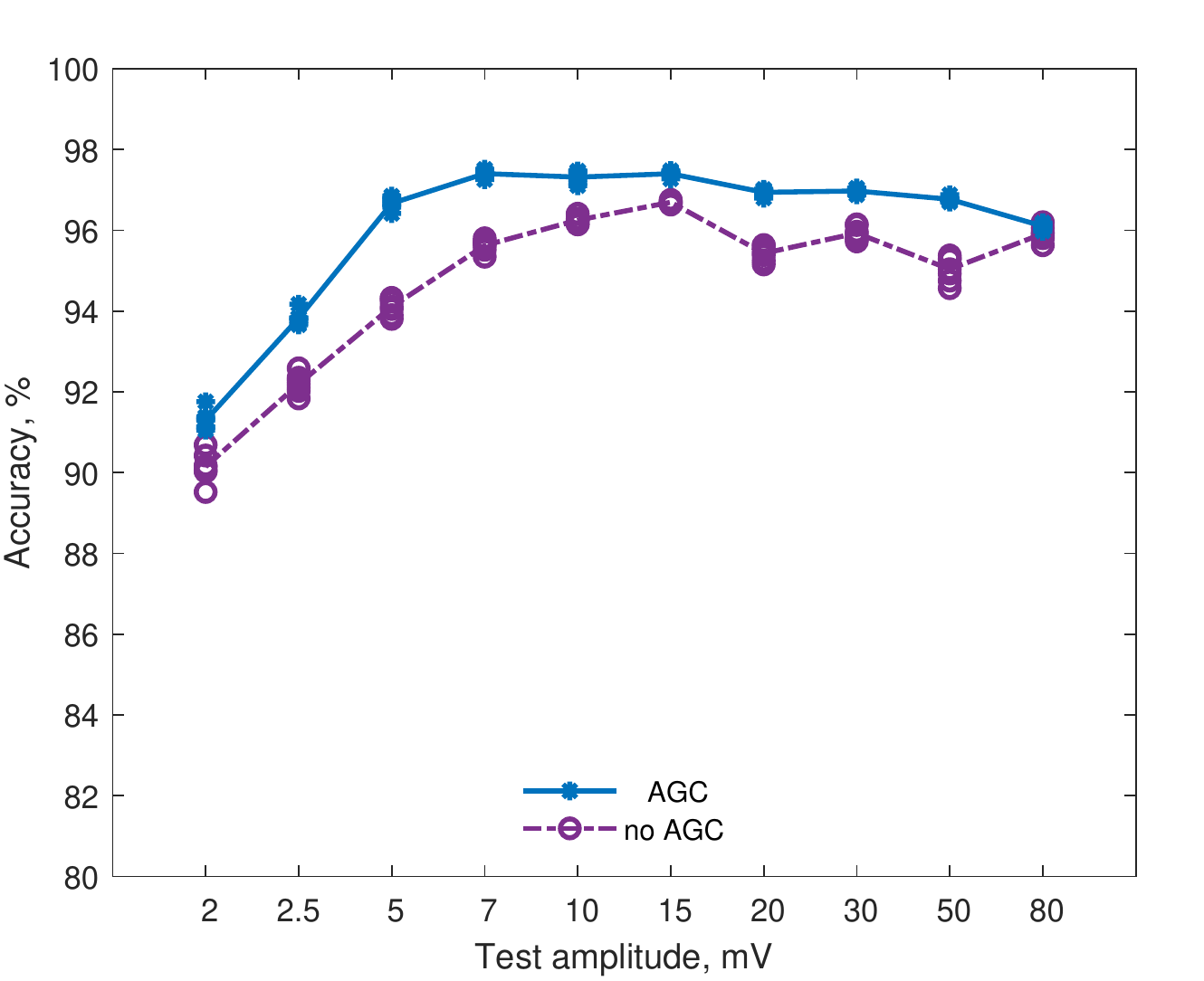}
\caption{DNN trained on all amplitudes in the training set.
}
\label{fig:exp_feat_all}}
\end{subfigure}

\caption{Speech vs noise accuracy of a 2L-64H fully-connected DNN trained on recordings at (a) $15$~mV$_{\textrm{RMS}}$ input and (b) all amplitudes. The classifier was tested across all amplitudes in the test set. 
}
\label{fig:exp_dnn}
\end{figure}

 \paragraph{LR classifier}

Figs.~\ref{fig:vad_lr} (a) and (b)
show the accuracy of the classifiers trained on the features acquired at two different input amplitudes ($15$~mV$_{\textrm{RMS}}$, $50$~mV$_{\textrm{RMS}}$) respectively. 
The plots show that as expected, the test accuracy is highest at the same input amplitude used during training. 
However, when the test data is acquired at a higher or lower input amplitude, the accuracy drop of the classifier for the non-AGC case is much higher than for the AGC case. 
The classifier trained for the AGC case achieves higher accuracy than one trained for the non-AGC case, when tested over all input amplitudes (Fig.~\ref{fig:vadall}). The average relative decrease in error of the models trained at different input amplitudes and tested over the full range of test amplitudes is shown in Fig.~\ref{fig:vad_error_decrease}. When the training and test input amplitudes differ by just $3$~dB, enabling AGC in the cochlea helps to decrease the relative error by about $10$\%.  The relative decrease in error increases up to $40$\% when the difference between the training and test signal amplitudes increases to 33 dB.

\begin{figure}[!t]
\centering
\includegraphics[width=0.45\textwidth]{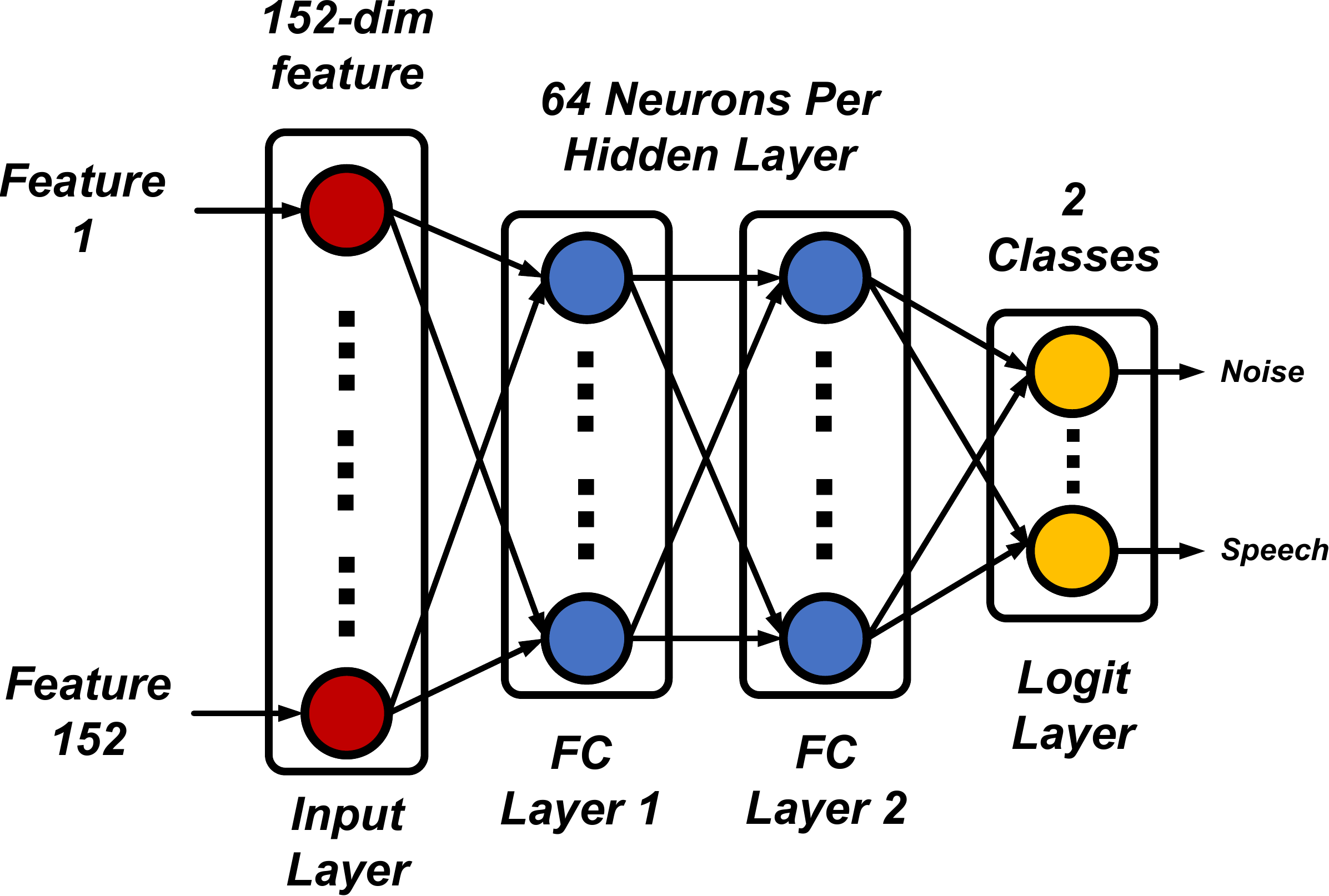}
\caption{The architecture of the DNN classifier.}
\label{fig:dnn}
\end{figure}

\paragraph{DNN classifier} 
We also tested a 2-layer fully-connected DNN with 64 neurons per layer on the same task. 
The network architecture is shown in Fig.~\ref{fig:dnn}. 
During training, we used the cross-entropy loss function and the ADAM optimizer. 
The network was trained for 50 epochs using a batch size of 64 and a learning rate of 5e-5. The dropout regularizer was applied to each FC layer after the ReLU activation functions with a dropout rate of 0.3.
A quarter of the training set was split into a validation set for early stopping at the highest accuracy on the validation set. Each dimension of the 152-dimensional feature vector was also normalized separately using the mean and the standard deviation of the dimension. The results reported were averaged over 6 runs. 
Fig.~\ref{fig:exp_dnn}(a) shows the test accuracy of the network trained with $15$~mV$_{\textrm{RMS}}$ input. Similar to the LR classifier results shown in Fig.~\ref{fig:vad_lr}(a), the classifier achieves the best test accuracy at the amplitude used during training. Overall, the AGC case outperforms the non-AGC case with only one exception at $80$~mV$_{\textrm{rms}}$. The degradation
of test accuracy becomes higher in both cases, at very low ($2$, $2.5$~mV$_{\textrm{RMS}}$) or very high test amplitudes ($50$, $80$~mV$_{\textrm{RMS}}$).

Fig.~\ref
{fig:exp_dnn}(b) shows
the accuracy of the DNN model trained on recordings of all amplitudes. Similar to the results in Fig.~\ref{fig:vad_lr}(c) accuracy improved across all test amplitudes compared to the case when the classifier was only trained on a single amplitude level. The DNN classifier outperforms the LR classifier reported in Fig.~\ref{fig:vad_lr}(c) as seen by the higher mean accuracy at all test amplitudes, and which are further summarized in Table~\ref{tab:acc}. Results from this table show that the non-AGC spike features benefit more from the DNN model than the AGC spike features across all trained conditions but the best accuracy from either  non-AGC or AGC spikes is higher with the DNN when compared to the accuracy from the logistic regression. To verify that the results were not affected by the dataset split, we did a 50:50 split for cross validation with 6 different random seeds on the amplitudes that exist in both the training set and the test set, including $5$, $10$, $15$, $50$, $80$~mV$_{\textrm{RMS}}$. Training on all amplitudes on the training set and testing on all amplitudes on the test set gave 96.24\% accuracy. Training on the test set and testing on the training set gave 96.18\%. The difference between the two results was only 0.06\%, showing that the dataset split has a minor impact on the accuracy.

The DNN can be deployed on an FPGA platform and would incur only a latency of around 10 ns~\cite{gaoDigit2019,gao2020edgedrnn}.  However, we did not implement this DNN on the FPGA because the work was to evaluate the classification accuracy of both the DNN and the logistic regression for the AGC and non-AGC cases.

\begin{table}[!t]
\caption{Mean test accuracy of the LR and DNN classifiers on  all test amplitudes. Results are averaged over 6 runs.}
\centering
\begin{tabular}{|c|c|c|c|c|}
\hline
\multirow{3}{*}{\textbf{\begin{tabular}[c]{@{}c@{}}Training \\ Amplitude \big(mV$_{\textrm{RMS}}$\big)\end{tabular}}} & \multicolumn{4}{c|}{\textbf{Mean Test Accuracy (\%)}}                \\ \cline{2-5}  & \multicolumn{2}{c|}{\textbf{LR}} & \multicolumn{2}{c|}{\textbf{DNN}} \\ \cline{2-5} & \textbf{Non-AGC}  & \textbf{AGC} & \textbf{Non-AGC}  & \textbf{AGC}  \\ \hline
5   & 69.79 & 72.45  & 77.45  & 74.68         \\ \hline
10  & 77.98    & 82.38   & 84.16   & 84.88  \\ \hline
15  & 77.85   & 85.42   & 86.01  & 88.12  \\ \hline
50  & 71.77   & 84.64  & 85.39  & 91.31         \\ \hline
80  & 71.06 & 83.76   & 86.04   & 90.19    \\ \hline
all    & 85.25   & 91.28   & 94.73    & 96.07   \\ \hline
\end{tabular}
\label{tab:acc}
\end{table}

\section{Discussion}
\label{sec:discussion}

This work presents an event-driven  spike-based controller for implementing local gain control in a spiking cochlea. 
With this AGC mechanism, we can extend the linear input range of a DASLP channel using the available gain of 32.5\,dB (Sec.~\ref{sec:methods_daslp})
on top of the 22\,dB measured at a single gain setting (Sec.~\ref{ssec:agc_analog_amp}).
This scheme has some limitations which we describe next.

Because the control scheme uses a step-wise gain update, a certain number of time steps are needed to reach the desirable channel filter output amplitudes. After each gain update, some length of time is needed for measuring the spike responses of every channel. This leads to a longer equivalent attack and release times for the gain control.
With our current gain control scheme, the averaging period for the spike responses is the same regardless of whether the gain should be increased or decreased, therefore the attack and release time ratios cannot be controlled unlike commercial global AGC chips which allow the setting of this ratio.

There are two additional limitations that come from the ASIC cochlea implementation. First, since only one channel on the ASIC can be selected for the update of the filter parameters, the gain can only be updated for one channel at a time. 
Second, because of the bias current settling time for a gain update on the ASIC, it is not possible to apply the AGC on high-frequency channels. A redesign of the chip to decrease the settling time of the attenuation levels would allow the AGC to be applied to all channels.
This can be mitigated by increasing the averaging window length, but would lead to increased attack and release times at low-frequency channels. 

Other future studies include how an increased spike-rate averaging window and hence, increased attack and release times, affect the AGC performance. Increasing the time averaging window would help to improve the AGC loop stability but increased attack and release times may deteriorate the advantages of including AGC. 

Other possible control schemes include the use of measured interspike intervals instead of spike counts, adjusting the upper threshold for the spike rate, and using  individual  thresholds  for each channel. The threshold can also be varied dynamically in time, e.g. decreasing the threshold in an absence of a signal would lead to a faster AGC response to the signal onset, and increasing the upper threshold in a presence of the signal would eliminate unnecessary gain changes during abrupt but short surges of the signal amplitude.
Other spike-based controller schemes such as the spike-based PID controller used in controlling a motor system ~\cite{perez2013neuro}, e.g., using spikes from a Dynamic Vision Sensor~\cite{li2015design,delbruck2021feedback}; and other event-triggered controllers~\cite{miskowicz2017event} might also be possible AGC candidates.

\section{Conclusion}
\label{sec:conclusion}

We present a system-level spike-based  
local automatic gain control mechanism which increases the linear input range of the DASLP spiking cochlea by 32.5\,dB.

There are two advantages of this AGC mechanism. One, the step-wise update control only needs counters and  comparators 
which can be 
implemented cheaply using either analog or digital circuits in
 a future cochlea ASIC design.
Two, when AGC is enabled, the spike rate of the spiking cochlea is compressed by $4$x over a wide range of input amplitudes (Sec.~\ref{ssec:agc_SS_gain_resp}). 
This compression in the spike rate will incur a lower number of computes, and therefore, power consumption in the post-processing event-driven or SNN hardware~\cite{davies2018loihi,Tsai2016SpeechTN,indiveri2015neuromorphic, neil2014minitaur}. The power consumption is potentially reduced by 4x for the  layer receiving the input spikes.

We further demonstrate the advantages of the resulting AGC-enabled spike features using two different classifiers in a speech classification task.
The logistic regression classifier achieves an improvement of 6\% in  classification accuracy or 40.8\% relative 
decrease in error with the AGC-enabled spike features. The results show that the AGC-enabled cochlea output carries more information for the classification task. The DNN classifier achieved a best accuracy of 96\% compared to the best accuracy of 91\% from the logistic regression.

The DNN classifier 
can be mapped into an ASIC design targeted at audio edge applications, similar to the VAD chip that combines a spiking cochlea filter bank together with a multi-layer perceptron \cite{yang2019vad} or in recent compute-in-memory based DNN accelerators, e.g.~\cite{Lee2021}, to realize highly energy-efficient inference. 
The AGC logic blocks can also be integrated on-chip with the cochlea circuits. 
The AGC mechanism can be applied towards other frequency-selective sensor front ends~\cite{MandalRFCochlea2020} that generate spikes and audio processing event-driven or SNN hardware platforms, e.g \cite{kiselev2016event,Tsai2016SpeechTN,SeokKWS2021} for operation in natural environments.

\section{Acknowledgments}
We thank Kwantae Kim for his help in improving the paper, and the helpful feedback from the anonymous reviewers.
% references section
%*********************************************************************************************
\bibliographystyle{IEEEtran}
\bibliography{myref}

\vskip 0pt plus -1fil
\begin{IEEEbiography}[{\includegraphics[width=1in,height=1.25in,clip,keepaspectratio]{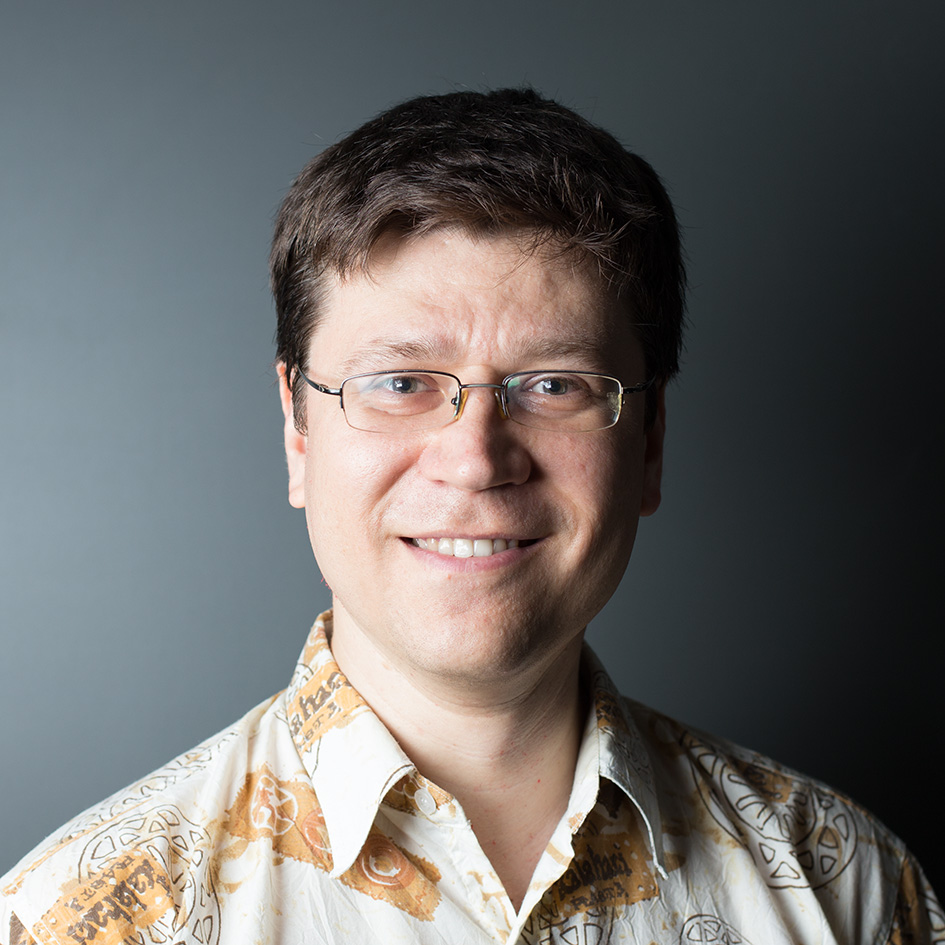}}]{Ilya Kiselev} (S'16) received his specialist degree in Physics at the Tambov State University (Russia) in 2000, his M.Sc. in Applied Mathematics and Physics from the Moscow Institute of Physics and Technology in 2002, and his doctoral degree from ETH Zurich in 2021. He is currently doing his postdoctoral work at the Institute of Neuroinformatics, University of Zurich and ETH Zurich. His research interests include hardware implementations of signal acquisition and processing for traditional and event-based audio processing. 
\end{IEEEbiography}
% \printbibliography
\vskip 0pt plus -1fil
\begin{IEEEbiography}[{\includegraphics[width=1in,height=1.25in,clip,keepaspectratio]{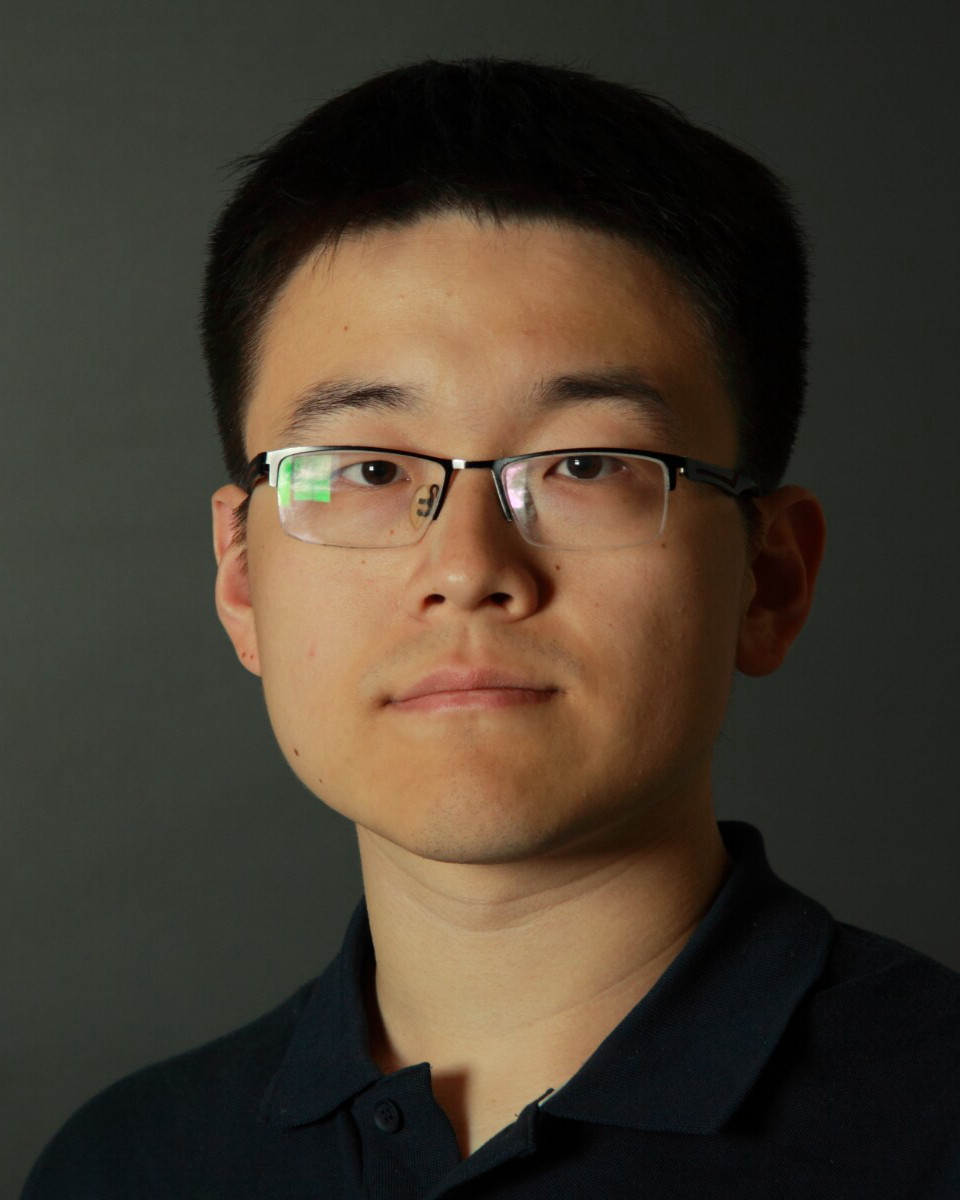}}]{Chang Gao}
(S’18–M’22) received his BEng degree in Electronics from University of Liverpool, Liverpool,
UK and Xi'an Jiaotong-Liverpool University in 2015, his MSc degree in Analog and Digital Integrated Circuit Design from Imperial College London in 2016, and his Ph.D. degree at the Institute of Neuroinformatics, University of Zurich and ETH Zurich in 2021. He is currently a postdoctoral researcher at the Institute of Neuroinformatics focusing on designing energy-efficient digital circuits for the acceleration of deep learning.
\end{IEEEbiography}
\vskip 0pt plus -1fil
\begin{IEEEbiography}[{\includegraphics[width=1in,height=1.25in,clip,keepaspectratio]{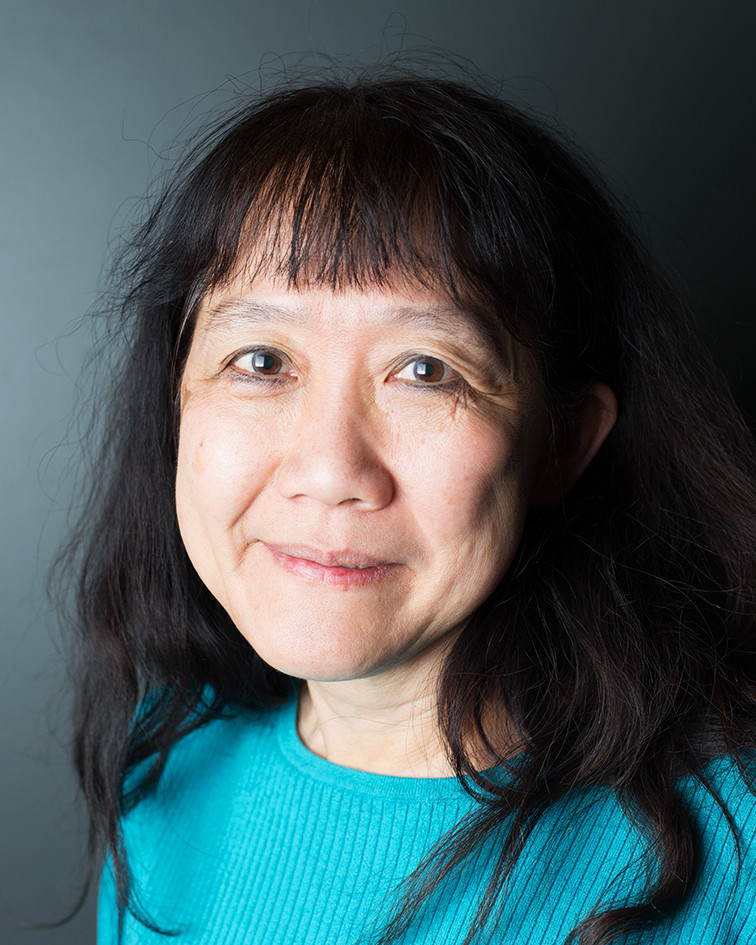}}]{Shih-Chii Liu}
(M’02–SM’07–F'22) received the bachelor’s degree in electrical engineering from the
Massachusetts Institute of Technology, Cambridge,
MA, USA, and the Ph.D. degree in the computation
and neural systems program from the California
Institute of Technology, Pasadena, CA, USA,
in 1997.
She is currently a Professor at the University of Zurich, Zurich Switzerland. Her group focuses on audio sensor designs, in particular, the spiking cochlea; and neuromorphic low-compute deep neural network algorithms and hardware.
\end{IEEEbiography}

\end{document}